\documentclass[
superscriptaddress,
twocolumn,
showpacs,preprintnumbers,
amsmath,amssymb,
aps,
prb,
]{revtex4-2}
\usepackage[colorlinks,
linkcolor=blue, urlcolor=blue, anchorcolor=blue, citecolor=blue]{hyperref}
\usepackage{graphicx}
\usepackage{dcolumn}
\usepackage{bm}
\usepackage{float}
\usepackage{multirow}
\usepackage{longtable}
\usepackage{amsmath}
\usepackage{amssymb}
\usepackage{xcolor}
\usepackage{lipsum}  
\usepackage[normalem]{ulem}
\usepackage{titlesec}

\newcommand{\angstrom}{\textup{\AA}}
\begin{document}


\title{Contrasting magnetism in VPS$_3$ and CrI$_3$ monolayers with the common honeycomb $S = 3/2$ spin lattice} 


\author{Ke Yang}
\affiliation{College of Science, University of Shanghai for Science and Technology, Shanghai 200093, China}
\affiliation{Laboratory for Computational Physical Sciences (MOE),
	State Key Laboratory of Surface Physics, and Department of Physics,
	Fudan University, Shanghai 200433, China}

\author{Yueyue Ning}
\affiliation{College of Science, University of Shanghai for Science and Technology, Shanghai 200093, China}

\author{Yaozhenghang Ma}
\affiliation{Laboratory for Computational Physical Sciences (MOE),
	State Key Laboratory of Surface Physics, and Department of Physics,
	Fudan University, Shanghai 200433, China}
\affiliation{Shanghai Qi Zhi Institute, Shanghai 200232, China}

\author{Yuxuan Zhou}
\affiliation{Laboratory for Computational Physical Sciences (MOE),
	State Key Laboratory of Surface Physics, and Department of Physics,
	Fudan University, Shanghai 200433, China}
\affiliation{Shanghai Qi Zhi Institute, Shanghai 200232, China}

\author{Hua Wu}
\email{Corresponding author. wuh@fudan.edu.cn}
\affiliation{Laboratory for Computational Physical Sciences (MOE),
	State Key Laboratory of Surface Physics, and Department of Physics,
	Fudan University, Shanghai 200433, China}
\affiliation{Shanghai Qi Zhi Institute, Shanghai 200232, China}
\affiliation{Hefei National Laboratory, Hefei 230088, China}

\date{\today}

\begin{abstract}

Two-dimensional (2D) magnetic materials are promising candidates for spintronics and quantum technologies. One extensively studied example is the ferromagnetic (FM) CrI$_3$ monolayer with the honeycomb Cr$^{3+}$ ($t_{2g}^3$, $S$ = 3/2) spin lattice, while VPS$_3$ has a same honeycomb $S$ = 3/2 spin lattice (V$^{2+}$, $t_{2g}^3$) but displays N$\acute{e}$el antiferromagnetism (AFM). In this work, we study the electronic structure and particularly the contrasting magnetism of VPS$_3$ and CrI$_3$ monolayers, using density functional calculations, magnetic exchange pictures, maximally localized Wannier functions (MLWFs) analyses, and parallel tempering Monte Carlo simulations. We find that VPS$_3$ is a Mott-Hubbard insulator but CrI$_3$ is a charge-transfer insulator, and therefore their magnetic exchange mechanisms are essentially different. The first nearest-neighbor (1NN) direct $d$-$d$ exchange dominates in VPS$_3$, thus leading to a strong antiferromagnetic (AF) coupling. However, the formation of vanadium vacancies, associated with instability of the low-valence V$^{2+}$ ions, suppresses the AF coupling and thus strongly reduces the N$\acute{e}$el temperature ($T_{\text{N}}$) in line with the experimental observation. In contrast, our results reveal that the major 1NN $d$-$p$-$d$ superexchanges in CrI$_3$ via different channels give rise to competing FM and AF couplings, ultimately resulting in a weak FM coupling as observed experimentally. After revisiting several important superexchange channels reported in the literature, based on our MLWFs and tight-binding analyses, we note that some antiphase contributions must be subtly and simultaneously considered, and thus we provide a deeper insight into the FM coupling of CrI$_3$. Moreover, we identify and compare the major contributions to the magnetic anisotropy, i.e., a weak shape anisotropy in VPS$_3$ and a relatively strong exchange anisotropy in CrI$_3$. Our work offers a comprehensive understanding of the 2D magnetism using the spin-orbital states and distinct exchange channels.
\end{abstract}

\maketitle


\section*{I. Introduction}

Since the discovery of two-dimensional (2D) ferromagnetic (FM) insulators CrI$_3$ and Cr$_2$Ge$_2$Te$_6$~\cite{Huang2017,Gong2017}, extensive research has focused on exploring the magnetic properties of 2D materials. According to the Mermin-Wagner theorem, magnetic anisotropy (MA) is essential for stabilizing long-range magnetic order in 2D materials~\cite{Mermin1966}, and therefore, quite a lot of studies were performed to seek the origin of MA~\cite{Lado2017,Kim2019,Yang2020,Zhao2021}. As more 2D materials have been discovered, increasingly complex magnetic structures have been observed, extending beyond FM ordering. Examples include N$\acute{e}$el antiferromagnetic (AF) ordering in MnPSe$_3$~\cite{Ni2021} and MnPS$_3$~\cite{Chu_2020}, zigzag AF ordering in FePS$_3$~\cite{Lee2016} and NiPS$_3$~\cite{Kang2020}, and even helimagnetic structures in NiI$_2$~\cite{Ju2021} and CoI$_2$~\cite{kim2023}.
Thus, both the magnetic couplings and MA are worthy of study for understanding of the diverse 2D magnetic structures.
\begin{figure}[t]
	\centering 
	\includegraphics[width=6cm]{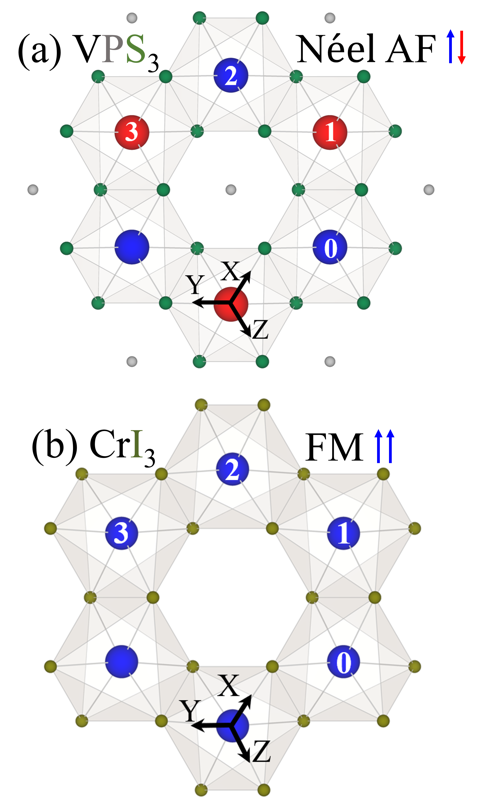}
	\centering
	\caption{The atomic structure of (a) N$\acute{e}$el AF VPS$_3$ and (b) FM CrI$_3$ monolayers. The blue and red spheres represent V/Cr ions with up and down spins, respectively.}
	\label{structure}
\end{figure}

Recently, the bulk van der Waals (vdW) material VPS$_3$ has been experimentally synthesized and exhibits out-of-plane N$\acute{e}$el AF insulating behavior with the N$\acute{e}$el temperature ($T_{\rm N}$) of 56 K~\cite{Liu2023}. The out-of-plane magnetic anisotropy and N$\acute{e}$el AF order persist even in thin samples, and the $T_{\rm N}$ of thin samples remains consistent with that of bulk VPS$_3$~\cite{Liu2023}.
Bulk VPS$_3$ is composed of stacked VPS$_3$ monolayers, each of which forms a honeycomb lattice structure similar to that of the extensively studied CrI$_3$~\cite{Huang2017}, see Fig.~\ref{structure}. In VPS$_3$, the V ions are in the 2+ valence state as in other $M$PS$_3$ compounds ($M$ = Mn~\cite{Chu_2020}, Fe~\cite{Lee2016}, Co~\cite{Wildes2017}, Ni~\cite{Kang2020} all in 2+ state), and each pair of P ions forms a dimer and each P ion is in the formal +4 state. Thus, the V$^{2+}$ ion in the local VS$_6$ octahedron adopts the $3d^3$ $t_{2g}^3$ ($S = 3/2$) configuration, being the same as the $t_{2g}^3$ Cr$^{3+}$ ion in CrI$_3$~\cite{Huang2017}. It is now surprising that although VPS$_3$ and CrI$_3$ have the common $S$ = 3/2 honeycomb lattice, the former is AF but the latter is FM, exhibiting contrasting magnetism. 
Moreover, previous theoretical studies have predicted that monolayer VPS$_3$ would exhibit N$\acute{e}$el AF order with a high $T_{\rm N}$ of 530 K~\cite{Bazazzadeh2021} or 570 K~\cite{Chittari2016}, which vastly deviates from the experimental bulk $T_{\rm N}$ of 56 K~\cite{Liu2023} as normally, the magnetic order temperature of a monolayer should be lower than that of its bulk.
Thus, we are motivated to investigate the origin of the contrasting magnetism between the $S$ = 3/2 honeycomb monolayers VPS$_3$ and CrI$_3$, and to reconcile the discrepancy between the theoretically predicted $T_{\rm N}$ and the experimental observation.

Here, we present a comparative study of the electronic structure and magnetic properties of VPS$_3$ and CrI$_3$ monolayers, using density functional theory (DFT) calculations, magnetic exchange pictures, maximally localized Wannier functions (MLWFs) analyses, and parallel tempering Monte Carlo (PTMC) simulations. Our results show that VPS$_3$ is a Mott-Hubbard insulator while CrI$_3$ is a charge-transfer insulator, and thus their magnetic exchange channels are essentially different. The major magnetic exchange in VPS$_3$ is the first nearest-neighbor (1NN) V-V direct exchange, and it gives rise to a strong AF coupling. However, the major one in CrI$_3$ is the Cr-I-Cr superexchange, and it leads to competing FM and AF exchange via different channels and to an ultimate weak FM coupling. Note that for the extensively studied CrI$_3$ monolayer, some antiphase contributions were overlooked and must now be subtly and simultaneously considered, after we revisited several leading exchange channels proposed in the literature. Moreover, we identify and compare major contributions of the magnetic anisotropy, $i.e.$, a weak shape anisotropy in VPS$_3$ and a relatively strong exchange anisotropy in CrI$_3$. Furthermore, to clarify the large discrepancy between the theoretical and experimental values of $T_{\rm N}$ for VPS$_3$, we demonstrate that the formation of vanadium vacancies in V$_{1-x}$PS$_3$ (against the less common low-valence V$^{2+}$ ions) remarkably weakens the otherwise strong AF coupling and then strongly reduces the theoretical $T_{\rm N}$. Then, the $T_{\rm N}$ value for VPS$_3$ and the $T_{\rm C}$ one for CrI$_3$ both obtained by our PTMC simulations are in agreement with the experimental values.

\section*{II. Computational Details}

Our DFT calculations are performed using the Vienna $Ab$ $initio$ Simulation Package (VASP)~\cite{Kresse1993}, with the generalized gradient approximation (GGA) as proposed by Perdew-Burke-Ernzerhof (PBE)~\cite{Perdew1996}. The optimized lattice constants of $a = b = 5.93$ $\angstrom$ for VPS$_3$ monolayer and $a = b = 6.99$ $\angstrom$ for CrI$_3$ are close to their respective experimental bulk values of 5.85~\cite{Klingen1970} and 6.87~\cite{McGuire2015} $\angstrom$. The total energies and atomic forces are converged to 10$^{-5}$ eV and 0.01 eV/$\angstrom$, respectively.
A 20 $\angstrom$ thick vacuum slab along the $c$-axis is used to model the VPS$_3$ and CrI$_3$ monolayers. The kinetic energy cutoff is set to 450 eV. For $k$-point sampling, a $\Gamma$-centered $k$-mesh of 9$\times$9$\times$1 is used for the 1$\times$1 unit cell, while a 9$\times$6$\times$1 mesh is used for the 1$\times$$\sqrt{3}$ supercell. The settings for the kinetic energy cutoff and $k$-point sampling have been carefully tested, concerning the subtle quantity of magnetic anisotropy energy (MAE) as detailed in Table S1 of the Supplemental Material (SM)~\cite{SM}.

To describe the on-site Coulomb interactions of V $3d$ and Cr 3$d$ electrons, we use the GGA + $U$ method~\cite{Anisimov1997} with a common Hubbard $U$ value of 4.0 eV and a Hund$'$s exchange parameter $J_{\rm H}$ of 0.9 eV~\cite{Yang2020,Huang_2018_jacs}. Moreover, hybrid functional calculations using the HSE06~\cite{HSE06} are carried out, and the obtained electronic structures are quite similar to the GGA + $U$ results, see Fig.~S1 in SM~\cite{SM}. 
Thus, the Mott-Hubbard insulating character of VPS$_3$ and the charge-transfer insulating behavior of CrI$_3$, as well as their consequent different magnetic exchanges, are found to be independent of the computational functional employed and are the primary focus of this work.
We also include the SOC effect in our GGA + SOC + $U$ calculations to study SOC-induced MA.
To study the magnetic exchange via different channels, the hopping parameters are obtained from MLWFs using the Wannier90 package~\cite{Mostofi2008, Nicola2012}. Additionally, we perform PTMC simulations~\cite{PTMC} to estimate the $T_{\rm N}$ of VPS$_3$ monolayer and the $T_{\rm C}$ of CrI$_3$ monolayer. These simulations are conducted on a 10$\times$10$\times$1 spin matrix, showing an insignificant change with respect to the tests also performed on 15$\times$15$\times$1 and 20$\times$20$\times$1 spin matrices, as seen in Fig.~S2 in SM~\cite{SM}. 
The number of replicas is set to 112.
During each simulation step, spins are randomly rotated in three-dimensional space, and the spin dynamics is analyzed using the classical Metropolis method~\cite{Metropolis1949}.

\section*{III. Results and Discussion}

\subsection*{A. VPS$_3$ monolayer: a Mott-Hubbard insulator}

\begin{figure}[t]
	\centering 
	\includegraphics[width=8.5cm]{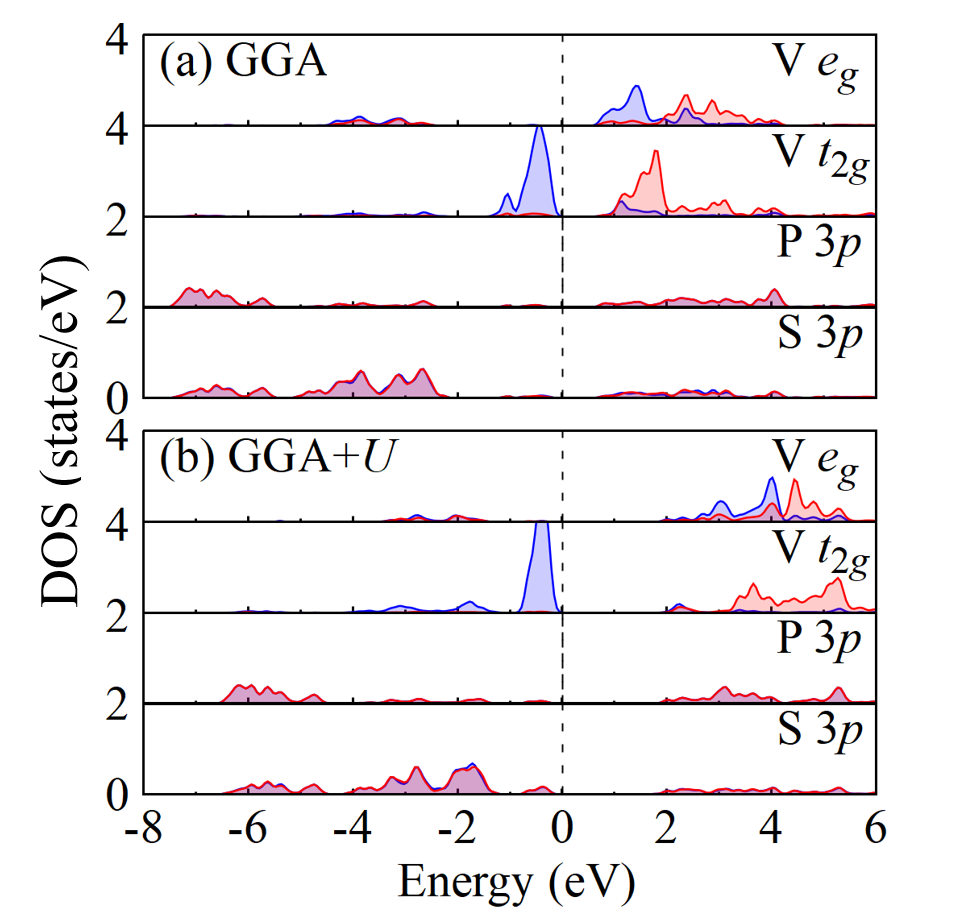}
	\centering
	\caption{Density of states (DOS) of VPS$_3$ monolayer in the N$\acute{e}$el AF state by (a) GGA and (b) GGA + $U$ calculations. The Fermi level is set at zero energy. The blue (red) curves stand for the up (down) spin channel.}
	\label{VPS3_dos}
\end{figure}

First, we investigate the electronic structure of the VPS$_3$ monolayer using GGA, and we plot in Fig.~\ref{VPS3_dos}(a) the orbitally resolved density of states (DOS) for the experimental N$\acute{e}$el AF state. 
The local octahedral coordination of VS$_6$ splits the five V $3d$ orbitals into a low-energy $t_{2g}$ triplet and a high-energy $e_g$ doublet.
Only the up-spin $t_{2g}$ triplet is fully occupied, giving the formal V$^{2+}$ $t_{2g}^3$ configuration with the total spin $S = 3/2$.
Both the valence bands and the conduction bands near the Fermi level are the V 3$d$ orbitals, and the bandwidths are each about 1 eV, being much smaller than the common Hubbard $U$ of several eV. This implies that the VPS$_3$ monolayer would be a Mott-Hubbard insulator. The S $3p$ bands lie below $-$2 eV relative to the Fermi level. The P $3p$ valence bands are much lower (below $-$6 eV), and some unoccupied P $3p$ components merge into the V $3d$ conduction bands, and the large bonding-antibonding splitting of about 8 eV is due to the P-P dimerization. The local spin moment of 2.15 $\mu_{\rm B}$ per V$^{2+}$ ion is reduced from the ideal 3 $\mu_{\rm B}$ due to the V $3d$-S $3p$ band hybridizations. 

For the strongly correlated VPS$_3$ with the narrow bands, our GGA + $U$ calculations show that the Mott gap is increased a lot by the Hubbard $U$ as shown in Fig.~\ref{VPS3_dos}(b). The electron correlation also increases the local spin moment up to 2.64 $\mu_{\rm B}$ per V$^{2+}$ ion, approaching the ideal 3 $\mu_{\rm B}$ as expected for the $S$ = 3/2.
Apparently, the first accessible electron excitation across the band gap in VPS$_3$ is the $d$-$d$ type, from the topmost V $3d$ valence bands to the lowest V $3d$ conduction bands, thus enabling us to classify VPS$_3$ as a Mott-Hubbard insulator.

\subsection*{B. The magnetic structure of VPS$_3$ monolayer}

\begin{figure}[t]
	\includegraphics[width=8cm]{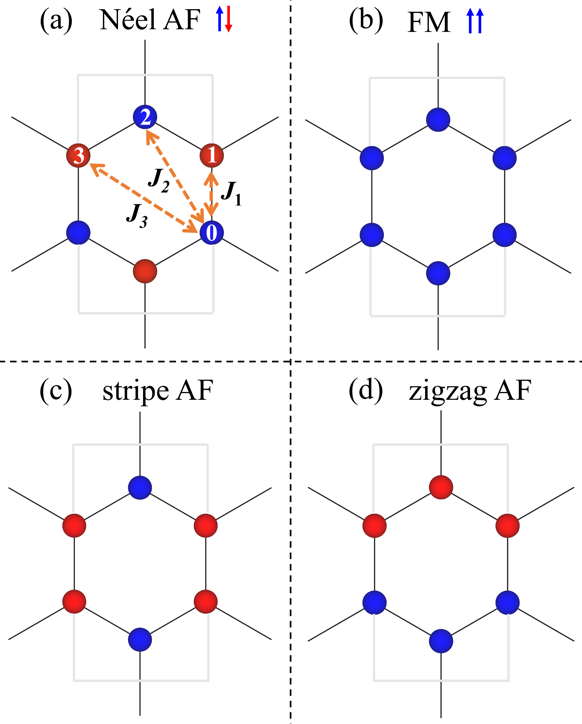}
	\centering
	\caption{The four magnetic structures of VPS$_3$ and CrI$_3$ monolayers, marked with three exchange parameters: 1NN ($J_1$), 2NN ($J_2$), and 3NN ($J_3$). The blue and red spheres represent V/Cr ions with up and down spins, respectively. Other atoms are hidden for simplicity.}
	\label{4Jstructure}
\end{figure}

To study the magnetic properties of the VPS$_3$ monolayer, we also perform GGA + $U$ calculations for other three magnetic structures (stripe AF, zigzag AF, and FM) in addition to the above N$\acute{e}$el AF state. All these magnetic structures are illustrated in Fig.~\ref{4Jstructure}, using a 1$\times\sqrt{3}$ supercell containing 4 V$^{2+}$ ions. Our results indicate that the N$\acute{e}$el AF solution is energetically most favorable, lying 52-164 meV/f.u. lower than the other three magnetic states, as shown in Table \ref{tb3J}. This result agrees with the experimental observation of the N$\acute{e}$el AF ground state~\cite{Liu2023}.
To understand these magnetic structures, we identify three exchange parameters: $J_1$ for 1NN V$_0$-V$_1$, $J_2$ for the 2NN V$_0$-V$_2$, and $J_3$ for the 3NN V$_0$-V$_3$, as seen in Fig.~\ref{4Jstructure}. 
This approach is often used in literature to deterimine the exchange parameters up to the 3NN for similar 2D magnetic materials~\cite{Sivadas_2015,Torelli_2018,Yang_2024}.
Considering the magnetic exchange energy $-JS^2$ ($J > 0$ for FM coupling) for each pair of V$^{2+}$ ions with $S = 3/2$, we express the relative exchange energies of the VPS$_3$ monolayer per formula unit for the four magnetic structures as follows:
\begin{equation}
	\begin{aligned}
		E_{\text{N$\acute{\rm e}$el AF}} &= (+\frac{3}{2}J_1-3J_2+\frac{3}{2}J_3)S^2 \\
		E_{\text{stripe AF}} &= (+\frac{1}{2}J_1+J_2-\frac{3}{2}J_3)S^2 \\
		E_{\text{zigzag AF}} &= (-\frac{1}{2}J_1+J_2+\frac{3}{2}J_3)S^2 \\
		E_{\text{FM}} &= (-\frac{3}{2}J_1-3J_2-\frac{3}{2}J_3)S^2 
	\end{aligned}
	\label{eq1}
\end{equation}
Using the relative total energies in Table \ref{tb3J} and applying Eq.~\ref{eq1}, we determine the exchange parameters for the VPS$_3$ monolayer to be $J_1 = -23.78$ meV, $J_2 = -0.56$ meV, and $J_3 = -0.52$ meV. These results show that all the three exchange couplings are AF, and the $J_1$ is about fifty times as big as the $J_2$ and $J_3$. Similar results are also obtained by the HSE06 calculations, see Table S2 in SM~\cite{SM}. Thus, the 1NN AF $J_1$ is indeed overwhelming. 

\renewcommand\arraystretch{1.3}
\begin{table}[t]
	\centering
	\caption{Relative total energies $\Delta$\textit{E} (meV/f.u.) and local spin moments ($\mu_{\rm B}$) for VPS$_3$ and CrI$_3$ monolayers obtained from GGA + $U$ calculations. The derived three exchange parameters (meV) are also listed.}
	\begin{tabular}{c@{\hskip8mm}r@{\hskip8mm}c@{\hskip8mm}r@{\hskip8mm}r@{\hskip8mm}}
		\hline\hline
		Systems	                        &  States             & $\Delta$\textit{E}  &  V/Cr   & S/I    \\ \hline
		\multirow{4}{*}{VPS$_3$}       & N$\acute{e}$el AF   &  0                  &  $\pm$2.64     & 0.00                          \\
		& stripe AF           &  52                 &  $\pm$2.66     & 0.00                          \\
		& zigzag AF           &  102                &  $\pm$2.68     & 0.00                            \\
		& FM                  &  164                &  2.72   & $-$0.01                                \\
		\multicolumn{5}{c}{\textit{J}$_{1}$ = $-$23.78 \hfill \textit{J}$_{2}$ = $-$0.56 \hfill \textit{J}$_{3}$ = $-$0.52}    \\ \hline
		\multirow{4}{*}{CrI$_3$}	    & FM                  &  0                  &  3.15                                & $-$0.09 \\
		& zigzag AF           &  9                  &  $\pm$3.12    & 0.00                             \\
		& stripe AF           &  15                 &  $\pm$3.11    & 0.00                       \\
		& N$\acute{e}$el AF   &  17                 &  $\pm$3.10    & 0.00                          \\
		\multicolumn{5}{c}{\textit{J}$_{1}$ = 2.56 \hfill \textit{J}$_{2}$ = 0.39 \hfill \textit{J}$_{3}$ = $-$0.04}    \\    	
		\hline\hline
	\end{tabular}
	\label{tb3J}
\end{table}

\subsection*{C. The origin of the N$\acute{e}$el AFM}

To understand the strong 1NN AF and the much weaker 2NN and 3NN AF couplings, we perform a Wannier function analysis to investigate the relevant hopping parameters associated with the different magnetic exchange channels, as seen in Fig.~S3(a) and Fig.~S4 in SM~\cite{SM}. This analysis focuses on the V 3$d$-S 3$p$ hybrid orbitals near the Fermi level, inherently encompassing both direct $d$-$d$ and indirect $d$-$p$-$d$ hoppings. Using these Wannier functions, we extract the hopping parameters between various orbitals of different V ions.

We first investigate the major hopping channels responsible for the strong 1NN AF coupling in the VPS$_3$ monolayer, where the V$_0$-V$_1$ bond distance is 3.42~$\angstrom$. As shown in Table~\ref{tbhopping_V}, diagonal hopping parameters within the two same $d$ orbitals are much larger than off-diagonal ones between two different $d$ orbitals, implying that direct $d$-$d$ hoppings between V-V ions may dominate. For the $t_{2g}^3$ V$^{2+}$ ions, the $e_g$ orbitals are empty. As a result, sizable hopping integrals between two empty $e_g$ orbitals, such as those between the two $3Z^2-R^2$ orbitals (92 meV) and between the two $X^2-Y^2$ orbitals (104 meV), do not contribute to magnetic coupling. Moreover, the hopping integrals between occupied $t_{2g}$ orbitals and unoccupied $e_g$ orbitals are much smaller than those between the diagonal occupied $t_{2g}$ orbitals. Therefore, we propose that the strong 1NN AF coupling in the VPS$_3$ monolayer primarily arise from the direct $d$-$d$ hoppings between the occupied $t_{2g}$ orbitals as demonstrated below.

\renewcommand\arraystretch{1.3}
\begin{table}[t]
	\centering
	\caption{The hopping parameters (meV) of VPS$_3$ monolayer.}
	\begin{tabular}{c@{\hskip3mm}l@{\hskip3mm}r@{\hskip3mm}r@{\hskip3mm}r@{\hskip3mm}r@{\hskip3mm}r@{\hskip3mm}} \hline\hline
		\multicolumn{2}{c}{\multirow{2}{*}{Hopping (\textit{t})}}  & \multicolumn{5}{c}{V$_0$} \\		      
		& & $3Z^2-R^2$ & $X^2-Y^2$ & $XY$ & $XZ$ & $YZ$ \\ \hline
		\multirow{5}{*}{V$_1$} & $3Z^2-R^2$ & $-$92 & $-$10 & $-$31 & $-$37 & 32 \\
		& $X^2-Y^2$ & $-$10 & $-$104 & 19 & $-$64 & $-$18 \\
		& $XY$ & $-$31 & 19 & 144 & 37 & $-$90 \\
		& $XZ$ & $-$37 & $-$64 & 37 & $-$432 & 37 \\
		& $YZ$ & 32 & $-$18 & $-$90 & 37 & 144 \\  \hline 
		\multirow{5}{*}{V$_2$} & $3Z^2-R^2$ & 13 & $-$23 & 74 & $-$47 & 0 \\
		& $X^2-Y^2$ & 23 & 24 & 16 & 25 & $-$44 \\
		& $XY$ & 75 & $-$16 & 2 & 25 & $-$11 \\
		& $XZ$ & 0 & 44 & $-$11 & 34 & $-$9 \\
		& $YZ$ & $-$48 & $-$25 & 25 & $-$34 & 34 \\ \hline
		\multirow{5}{*}{V$_3$} & $3Z^2-R^2$ & 150 & 103 & 28 & $-$8 & $-$57 \\
		& $X^2-Y^2$ & 103 & 32 & $-$7 & $-$28 & 98 \\
		& $XY$ & 28 & $-$7 & 6 & $-$18 & 0 \\
		& $XZ$ & $-$8 & $-$28 & $-$18 & 6 & 1 \\
		& $YZ$ & $-$57 & 98 & 0 & 1 & $-$106 \\ 
		\hline\hline
	\end{tabular}
	\label{tbhopping_V}
\end{table}

\begin{figure}[t]
	\centering 
	\includegraphics[width=8.5cm]{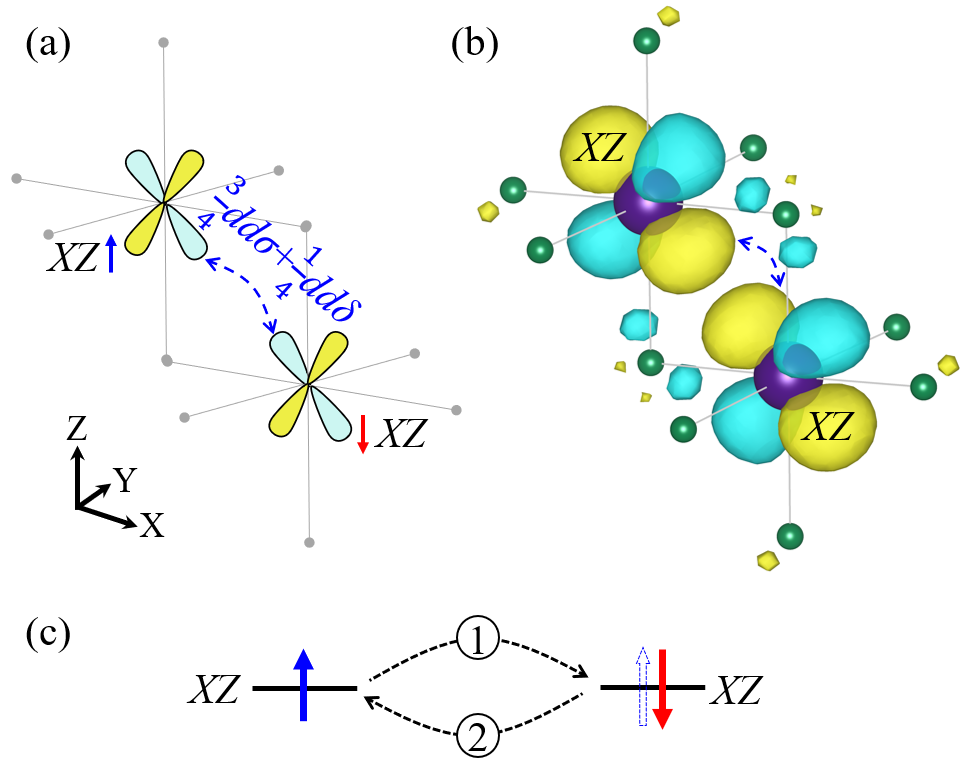}
	\centering
	\caption{(a) The hybridization of the two $XZ$ orbitals and (b) the corresponding Wannier orbitals in VPS$_3$. (c) The direct $d$-$d$ hopping channels between the two $XZ$ orbitals lead to AF coupling.}
	\label{VPS3_xz_xz_1}
\end{figure}

As seen in Table~\ref{tbhopping_V}, the largest 1NN hopping integral (432 meV) occurs between the two $XZ$ orbitals, while other hopping integrals are all less than one-third of this value. To understand why such a large hopping integral exists between the two $XZ$ orbitals, we illustrate the real-space distribution of the $XZ$-like MLWFs in Fig.~\ref{VPS3_xz_xz_1}(b). In the edge-sharing octahedra, the two $XZ$ orbitals on adjacent V$_0$-V$_1$ sites are aligned toward each other, enabling a direct $d$-$d$ hopping with the integral of $\frac{3}{4}$$dd\sigma$ + $\frac{1}{4}$$dd\delta$, as shown in Fig.~\ref{VPS3_xz_xz_1}(a). This would yield a strong AF coupling, considering the Pauli exclusion principle after the $d$-$d$ hopping. 
When considering the indirect $d$-$p$-$d$ hopping channel between the two $XZ$ orbitals via the intermediate S ligand$'$s $p$ orbitals, it is important to note that a single $p_X$ or $p_Z$ orbital of the S ion is nearly orthogonal to one of the two $XZ$ orbitals (see Table S3 in the SM~\cite{SM}), thus rendering the indirect $d$-$p$-$d$ hopping channel almost ineffective.
Therefore, a superexchange can only occur simultaneously through the $p_X$ and $p_Z$ orbitals of the intermediate S ion, as shown in Fig.~S5 in SM~\cite{SM}. In this case, a virtual charge fluctuation corresponds to a transition from the ground state $3d^3-3p^6-3d^3$ to an excited state $3d^4-3p^4-3d^4$ with the double holes $p_xp_z$ on the intermediate S ion. Given that VPS$_3$ is a Mott-Hubbard insulator preferring the $d-d$ excitation, the energy cost to the excited intermediate $3d^4-3p^4-3d^4$ state is much larger than the direct $d-d$ excitation into $3d^2-3d^4$. Thus, the superexchange associated with the virtual excitation $3d^4-3p^4-3d^4$ would yield a much weaker FM coupling.
To further verify that the large hopping integral between the two $XZ$ orbitals originates from the direct $d$-$d$ hybridization, we conducted a separate Wannier function analysis, projecting onto both the V 3$d$ and S 3$p$ orbitals, as seen in Fig.~S3(b) and Fig.~S6 in SM~\cite{SM}. In this calculation, the hopping integral between the two $XZ$ orbitals accounts exclusively for the direct $d$-$d$ hybridization, and our results reveal the hopping integral as large as 341 meV, as seen in Table S4 in the SM~\cite{SM}, further confirming this conclusion.
Thus, as shown in Fig.~\ref{VPS3_xz_xz_1}(c), when considering virtual charge fluctuations, local Hund$'$s exchange, and the Pauli exclusion principle, the dominant direct $d$-$d$ hopping channels between the two $XZ$ orbitals lead to the strong 1NN AF coupling.

\begin{figure}[t]
	\centering 
	\includegraphics[width=8.5cm]{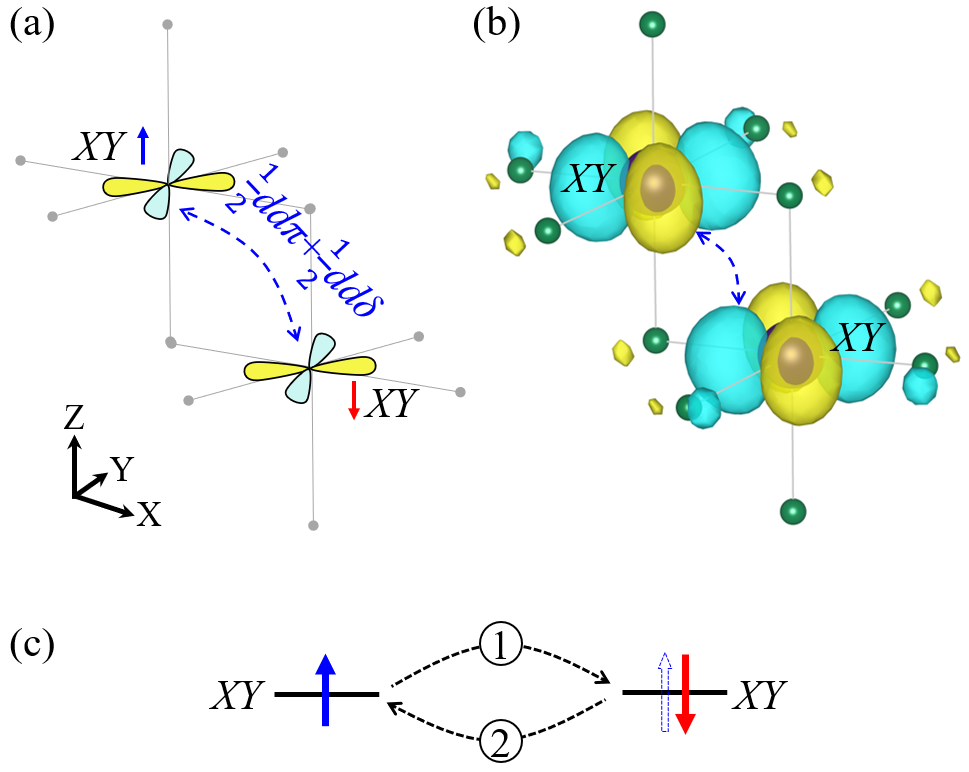}
	\centering
	\caption{(a) The hybridization of the two $XY$ orbitals and (b) the corresponding Wannier orbitals in VPS$_3$. (c) The direct $d$-$d$ hopping channels between the two $XY$ orbitals lead to AF coupling.}
	\label{VPS3_xy_xy_1}
\end{figure}

The second largest 1NN hopping integral (144 meV) arises between two $XY$ orbitals and between two $YZ$ orbitals. As shown in Fig.~\ref{VPS3_xy_xy_1} and Fig.~S7 in SM~\cite{SM}, the two $XY$ or two $YZ$ orbitals are aligned in parallel, both providing the same direct hopping integral of $\frac{1}{2}dd\pi+\frac{1}{2}dd \delta$. Moreover, since the intermediate S $p_X$, $p_Y$, or $p_Z$ orbital is each orthogonal to at least one of the two $XY$ (or one of the two $YZ$) orbitals, the indirect $d$-$p$-$d$ hopping integral between the two $XY$ or between two $YZ$ orbitals is zero. As shown in Fig.~\ref{VPS3_xy_xy_1}(c) and Fig.~S7(c) in SM~\cite{SM}, the direct $d$-$d$ hopping channels between the two $XY$ and between two $YZ$ orbitals both result in an AF coupling.

\begin{figure}[b]
	\centering 
	\includegraphics[width=8.5cm]{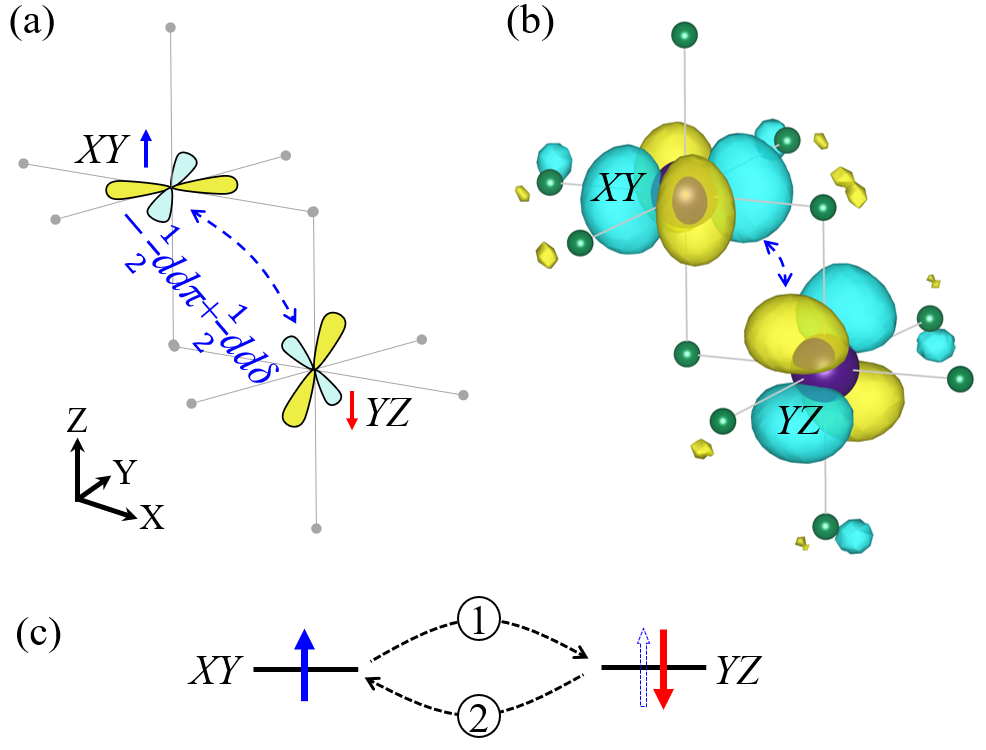}
	\centering
	\caption{(a) The hybridization bewteen the $XY$ and $YZ$ orbitals and (b) the corresponding Wannier orbitals in VPS$_3$. (c) The direct $d$-$d$ hopping channels between the $XY$ and $YZ$ orbitals lead to AF coupling.}
	\label{VPS3_xy_yz_1}
\end{figure}

Then, we analyze the off-diagonal 1NN hopping integral of 90 meV between the $XY$ and $YZ$ orbitals. As shown in Fig.~\ref{VPS3_xy_yz_1}, the direct $d$-$d$ hopping with the integral of $-\frac{1}{2}$$dd\pi$ + $\frac{1}{2}$$dd\delta$ also leads to an AF coupling. In addition, the ($XY$)-($p_Y$)-($YZ$) superexchange channel exists (see Fig.~S8) and it also contributes to an AF coupling. Thus, from the above results and analyses, we can propose that in the VPS$_3$ monolayer, the strong 1NN AF $J_1 = -23.78$ meV primarily arises from the direct $d$-$d$ hoppings/exchanges between the $t_{2g}$ orbitals on the 1NN V$^{2+}$-V$^{2+}$ pairs.


In line with the above computations of $J_2 = -0.56$ meV and $J_3 = -0.52$ meV, both being much smaller in strength than $J_1 = -23.78$ meV, the hopping parameters for the 2NN V$_0$-V$_2$ and 3NN V$_0$-V$_3$ are much smaller than those for the 1NN V$_0$-V$_1$, as shown in Table~\ref{tbhopping_V}. This comparison is also rationalized by an estimate of the exchange parameters which are proportional to $t^2/U$ where $t$ is the different hopping parameters and $U$ is the on-site Coulomb repulsion of the V $3d$ electrons. It is now a bit surprising that why $J_3$ is well comparable to $J_2$, considering the larger 3NN V-V ion distance of 6.85 $\angstrom$ than the 2NN distance of 5.93 $\angstrom$. As the direct $d$-$d$ hopping is negligible for the 2NN and 3NN, the indirect hopping channels should play the major role. Note that for $J_3$, there is an additional superexchange channel mediated by the fat orbitals of the P-P dimer at the center of the honeycomb lattice, and then this channel provides a large hopping parameter of 106 meV between two occupied $YZ$ orbitals, see Fig. S9 in SM~\cite{SM}. Note also that although the unoccupied $e_g$ orbitals have large hopping integrals of 150 meV or 103 meV, they do not contribute to the magnetic couplings. Then, the hopping integral of 106 meV should be an important reason why $J_3$ is well comparable with $J_2$, and this point will also be implied in the following sections about CrI$_3$ monolayer where a void space appears at the center of the Cr$^{3+}$ honeycomb lattice and then $J_3$ is significantly suppressed, see $J_3 = -0.04$ meV vs $J_2 = 0.39$ meV in Table I.

\subsection*{D. The out-of-plane MA and high T$_N$}

With the above understanding of the magnetic couplings in the VPS$_3$ monolayer, we now turn our attention to the other crucial aspect of 2D magnetism, $i.e.$, the magnetic anisotropy. To calculate the MA of the VPS$_3$ monolayer, we perform GGA + SOC + $U$ calculations.
Our results indicate that the VPS$_3$ monolayer favors the out-of-plane magnetization being consistent with the experimental observations~\cite{Liu2023}, while the in-plane magnetization is slightly higher in energy by 12 $\mu$eV per V$^{2+}$ ion. This small magnetic anisotropy energy is attributed to the closed V$^{2+}$ $t_{2g}^3$ shell (with $S$ = 3/2 and $L$ = 0), which results in a negligible single ion anisotropy. Additionally, the exchange anisotropy is also weak, arising from the limited SOC of the S 3$p$ orbitals and their hybridization with V 3$d$ orbitals. In view of the small SOC-induced MAE, the shape anisotropy due to magnetic dipole-dipole interactions may be of concern.

We now consider the magnetic dipole-dipole interaction whose energy is expressed as
\begin{equation} E^{\rm dipole} = \frac{\mu_0}{4\pi} \frac{1}{r_{12}^3} \left[ \overrightarrow{M}_1 \cdot \overrightarrow{M}_2 - \frac{3}{r_{12}^2} (\overrightarrow{M}_1 \cdot \overrightarrow{r}_{12})(\overrightarrow{M}_2 \cdot \overrightarrow{r}_{12}) \right] \label{eq6} \end{equation}
where $\overrightarrow{M}_1$ and $\overrightarrow{M}_2$ are the magnetic moments of the two dipoles, and $\overrightarrow{r}_{12}$ is the vector connecting them.
When these two magnetic dipoles are aligned parallel (FM), the energy difference between in-plane and out-of-plane magnetizations is
\begin{equation} E_{\rm FM}^{\parallel} - E_{\rm FM}^{\perp} = -\frac{3\mu_0 M^2 \cos^2 \theta}{4\pi r_{12}^3} \label{eq9} \end{equation}
where $\theta$ is the angle between the $\overrightarrow{M}_1$ and $\overrightarrow{r}_{12}$. This expression shows that the shape anisotropy favors in-plane magnetization for the FM state.

In contrast, when the two dipoles are aligned antiparallel (AF), the energy difference is
\begin{equation} E_{AF}^{\parallel} - E_{AF}^{\perp} = \frac{3\mu_0 M^2 \cos^2 \theta}{4\pi r_{12}^3}. \label{eq12} \end{equation}
Then, the shape anisotropy favors out-of-plane magnetization for the AF state.

\begin{figure}[t]
	\centering 
	\includegraphics[width=8.5cm]{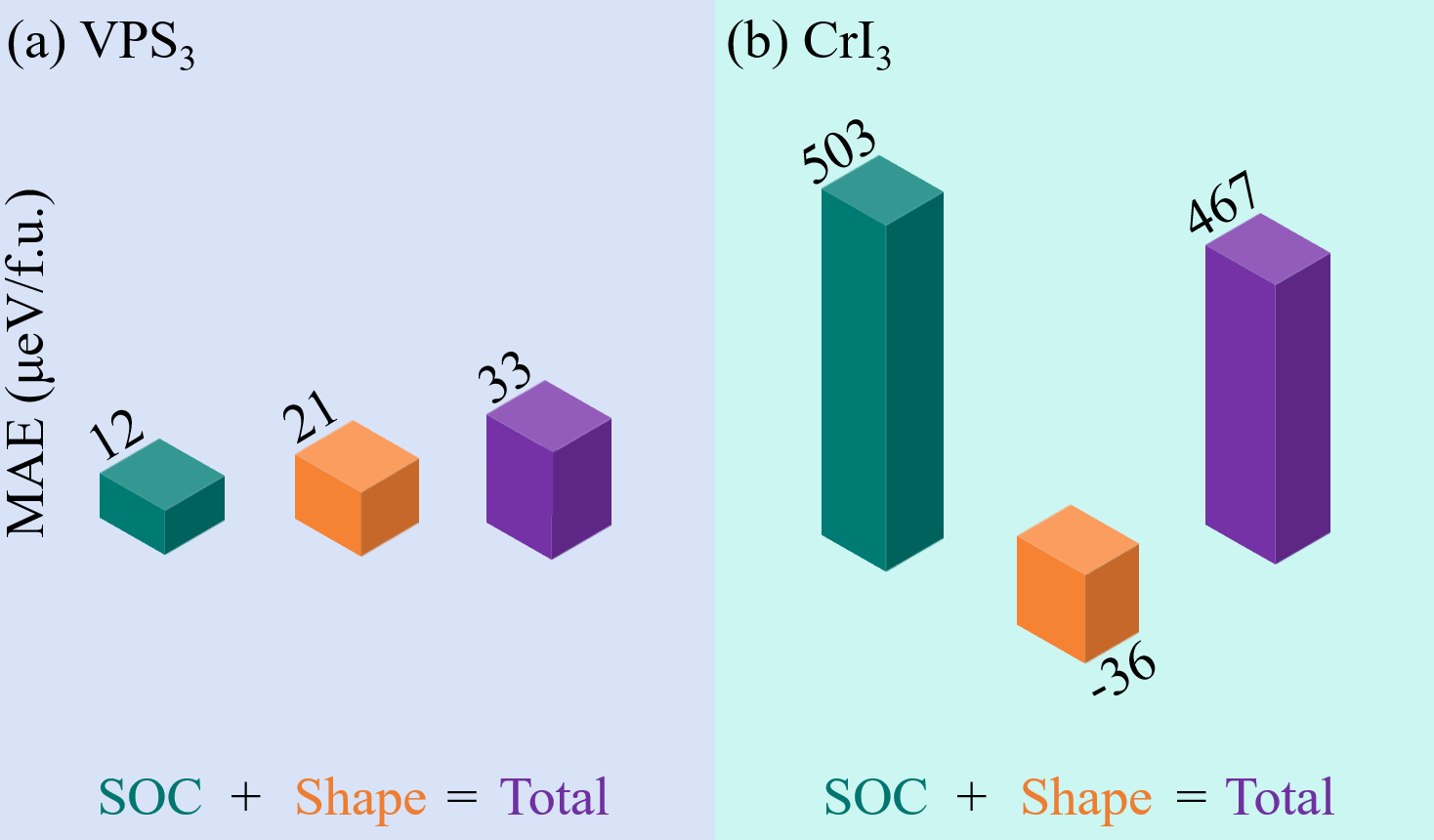}
	\centering
	\caption{The SOC-MAE, shape-MAE, and total-MAE ($\mu$eV/f.u.) for (a) VPS$_3$ in the  N$\acute{e}$el AF state and (b) CrI$_3$ in the FM state.}
	\label{MAE}
\end{figure}

For real materials, shape anisotropy is a long-range magnetic dipole-dipole interaction, which requires considering contributions from different nearest neighbors. For VPS$_3$ in the N$\acute{e}$el AF ground state [see Fig.~\ref{4Jstructure}(a)], the 1NN and 3NN AF couplings favor the out-of-plane shape anisotropy, while the 2NN FM coupling favors the in-plane shape anisotropy. Since the shape anisotropy is antiproportional to $r_{12}^{3}$, the N$\acute{e}$el AF VPS$_3$ would ultimately be expected to  exhibit the out-of-plane shape anisotropy.
We present the shape-MA results in Fig.~\ref{MAE}, indeed confirming the out-of-plane magnetization for the N$\acute{e}$el AF VPS$_3$. The in-plane magnetization has a higher shape MAE by 21 $\mu$eV per V$^{2+}$ ion, being nearly twice the SOC-induced MAE of 12 $\mu$eV per V$^{2+}$ ion. Therefore, the experimental easy out-of-plane magnetization in VPS$_3$ is a joint effect of the shape-MA and SOC-MA.

Using the calculated exchange parameters and the out-of-plane MAE, we assume a spin Hamiltonian and carry out PTMC simulations to estimate the $T_{\rm N}$ of VPS$_3$ monolayer
\begin{equation} \label{eq5}
	H=-\sum_{k = 1, 2, 3}\sum_{i,j} \frac{J_{k}}{2} \mathbf{S}_{i} \cdot \mathbf{S}_{j}-\sum_{i} D\left(S_{i}^{z}\right)^{2}.
\end{equation} 
The first term represents the isotropic Heisenberg exchange, summing over all the V$^{2+}$ sites $i$ (with $S = 3/2$) in the honeycomb spin lattice, while $j$ runs over the $k$NN V$^{2+}$ sites of each $i$, with their respective AF couplings $J_k$ given as $J_1 = -23.78$ meV, $J_2 = -0.56$ meV, and $J_3 = -0.52$ meV. The second term describes the out-of-plane MAE of 33 $\mu$eV per V$^{2+}$ ion, with $D$ = 0.015 meV. Then, our PTMC simulations yield the $T_{\rm N}$ of 276 K for the VPS$_3$ monolayer, as shown in the inset of Fig.~\ref{MC}. If the Ising limit is assumed as in Ref.~\cite{Chittari2016} where $T_{\rm N}$ was calculated to be 570 K, here the $T_{\rm N}$ would be increased up to 668 K. Note that all these $T_{\rm N}$ values are overestimated, compared with the experimental $T_{\rm N}$ of 56 K~\cite{Liu2023} for the bulk VPS$_3$ ($T_{\rm N}$ would be reduced for monolayer due to the dimensionality effect). As demonstrated below, this discrepancy is most probably due to the V vacancies.

\subsection*{E. Vanadium vacancy and the much reduced $T_{\rm N}$}
\begin{figure}[t]
	\centering 
	\includegraphics[width=8cm]{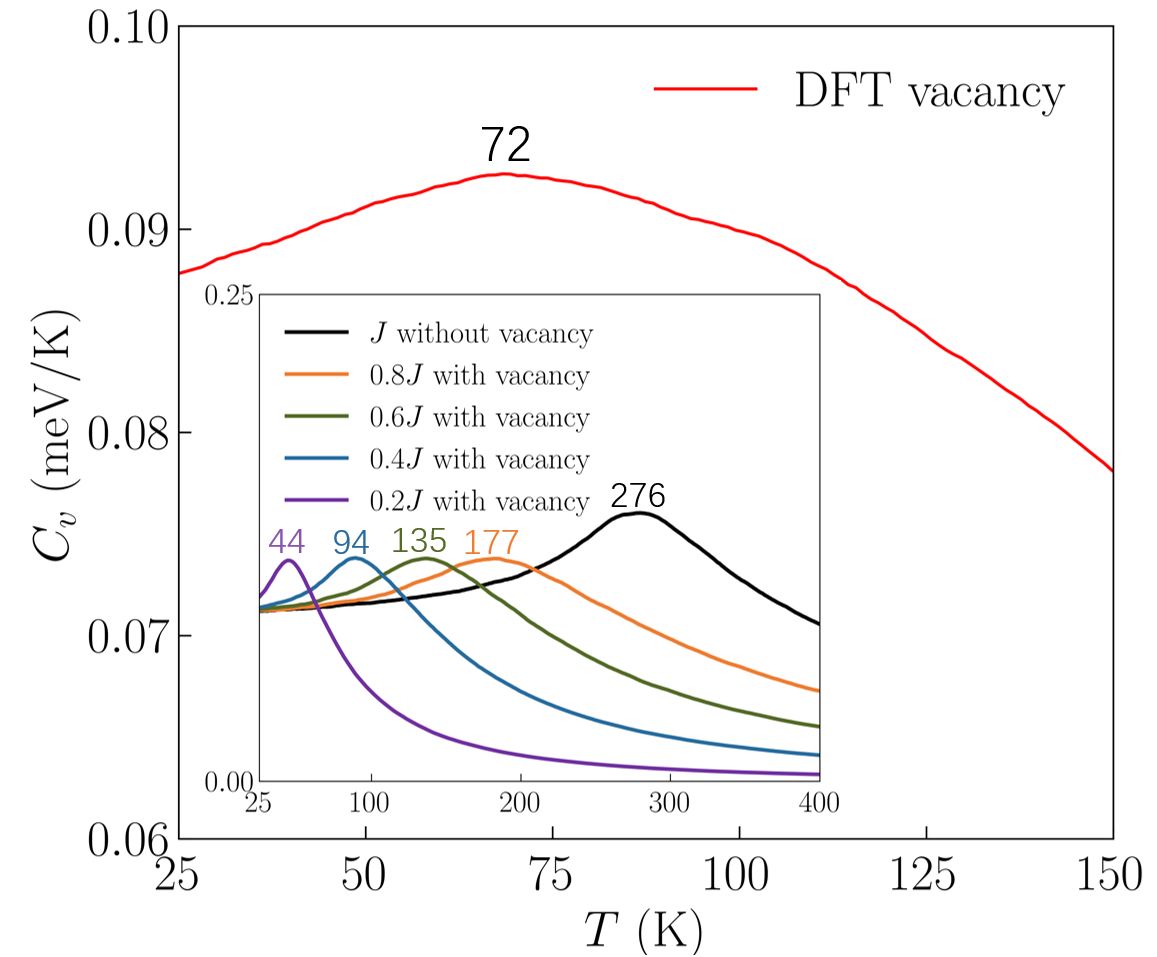}
	\centering
	\caption{PTMC simulations of the magnetic specific heat of the VPS$_3$ monolayer. The red curve represents the simulation using V vacancy-induced FM and reduced AF exchange parameters (calculated from DFT with V vacancy modeling), showing that $T_{\rm N}$ could be as low as 72 K. The inset black curve represents the $T_{\rm N}$ at 276 K without V vacancies. Other curves in the inset show our simulations in which 10\% V vacancies weaken the magnetic coupling, causing $T_{\rm N}$ to drop to 135 K (at 0.6$J$) and even further to 44 K (at 0.2$J$).}
	\label{MC}
\end{figure}

The V$^{2+}$ ions are less common (+3, +4, and +5 are common charge states for vanadium ions, $e.g.$, in V$_2$O$_3$, VO$_2$, V$_2$O$_5$, and Cr doped VO$_2$~\cite{Pandey_2021}, and some of them would be converted into the common V$^{3+}$ when the V vacancies appear, and this indeed occurs in the previous studies about V$_{1-x}$PS$_3$~\cite{Coak,Coak_2019,Coak1985}. Assuming a composition of V$_{0.9}$PS$_3$, the lattice distortion caused by the V vacancies would suppress the aforementioned AF couplings, and even some V$^{2+}$-V$^{3+}$ pairs probably yield a double exchange FM coupling. As such, the average AF coupling should be remarkably weakened and thus a significantly lowering $T_{\rm N}$ would be expected for V$_{0.9}$PS$_3$ monolayer. To check this effect, we performed PTMC simulations using a two-dimensional 10$\times$10$\times$1 spin lattice with 10\% V vacancies. Our results indicate that the $T_{\rm N}$ of 276 K would drop down to 135 K when the above three AF $J_k$ values are scaled down to 0.6$J_k$, and even further down to 44 K for 0.2$J_k$, as seen in the inset of Fig.~\ref{MC}. 

To provide a direct proof for the significant reduction of $T_{\rm N}$ by the V vacancies, we performed DFT calculations using a 2$\times$2 supercell with a single V vacancy (with the vacancy ratio of 1/8; see Fig.~S10 in SM~\cite{SM}). The supercell structure is relaxed using GGA, and owing to the V vacancy and the resulting lattice distortion, the inequivalent V sites emerge in the supercell and lead to the varying magnetic exchange parameters.
For the V site farthest from the vacancy, our GGA + $U$ calculations give an average 1NN AF exchange parameter of $-15.37$ meV (about 65\% of the above $J_1$ = -- 23.78 meV). However, for the V site closest to the vacancy, the average 1NN exchange parameter is 5.26 meV of the FM type. 
[Other exchange parameters are expected to lie in between them, and the average exchange (most likely AF) among all those various magnetic channels in the supercell with a V vacancy should be much weaker than that of the otherwise homogeneous lattice.] 
Considering a random distribution of the 10\% V vacancies in the 10$\times$10$\times$1 spin lattice, and assuming the ending limit AF or FM exchange dependent on the location of the V ions away from each vacancy, we performed PTMC simulations and found that $T_{\rm N}$ could be as low as 72 K and it should be even lower when using many other AF parameters weaker than the value of $-15.37$ meV. Such a $T_{\rm N}$ is quite close to the experimental 56 K~\cite{Liu2023}. Therefore, we could claim that the vanadium vacancies significantly reduce the $T_{\rm N}$ of VPS$_3$, and they well account for the discrepancy between the computed high $T_{\rm N}$ for the ideal/perfect structure and the experimental low $T_{\rm N}$ due to the vacancies. To achieve the ideal high $T_{\rm N}$ of VPS$_3$ monolayer, the V vacancies should be minimized during the sample preparation/growth.

\subsection*{F. CrI$_3$ monolayer: a charge-transfer insulator and FM superexchange}

After identifying the Mott-Hubbard insulating and AF character of VPS$_3$ monolayer, we now turn our attention to the intriguing CrI$_3$ monolayer~\cite{Lado2017,Kim2019,Zhao2021,McGuire2015,Huang2017,Xue2019} to understand why it has the contrasting FM order despite having the same honeycomb $S$ = 3/2 lattice as the AF VPS$_3$. As demonstrated below, the FM order in CrI$_3$ is closely related to the charge-transfer insulating behavior and the consequent FM superexchange, rather than the direct AF exchange in the Mott-Hubbard insulating VPS$_3$. Moreover, we identify the strongest FM superexchange channel by a detailed analysis of the hopping integrals via Wannier functions. Thus, we provide a new insight into the FM order of the extensively studied CrI$_3$ monolayer.

\begin{figure}[t]
	\centering 
	\includegraphics[width=8cm]{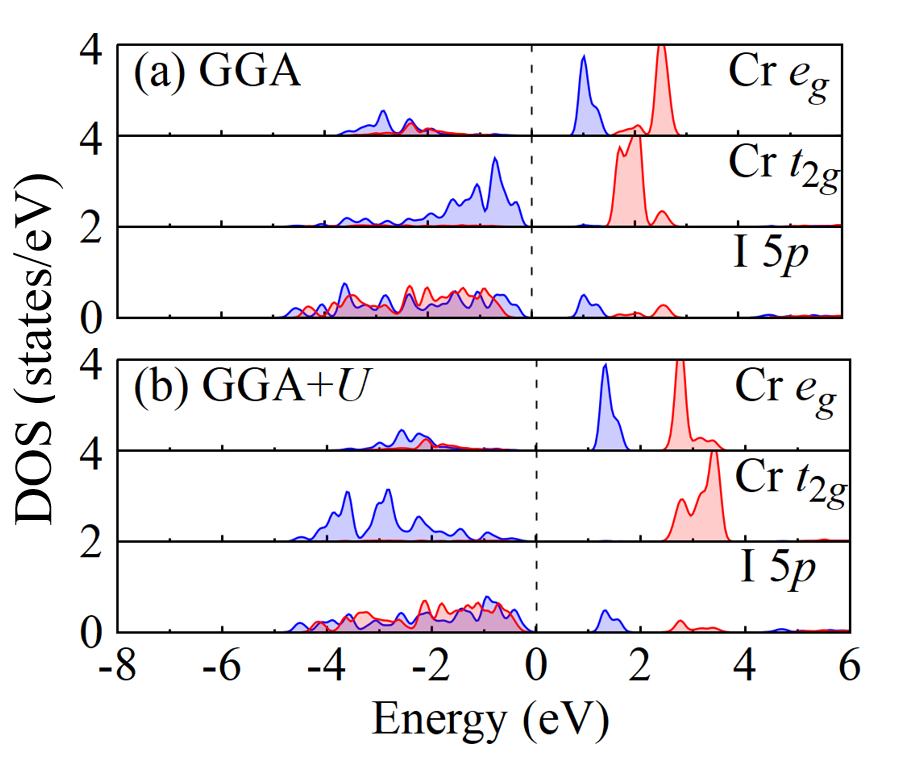}
	\centering
	\caption{Density of states (DOS) of CrI$_3$ monolayer by (a) GGA and (b) GGA + $U$ calculations. The Fermi level is set at zero energy. The blue (red) curves stand for the up (down) spin channel. 
	}
	\label{CrI3_dos}
\end{figure}
\renewcommand\arraystretch{1.3}
\begin{table}[b]
	\centering
	\caption{The hopping parameters (meV) of CrI$_3$ monolayer.}
	\begin{tabular}{c@{\hskip3mm}l@{\hskip3mm}r@{\hskip3mm}r@{\hskip3mm}r@{\hskip3mm}r@{\hskip3mm}r@{\hskip3mm}} \hline\hline
		\multicolumn{2}{c}{\multirow{2}{*}{Hopping (\textit{t})}}  & \multicolumn{5}{c}{Cr$_0$} \\		      
		& & $3Z^2-R^2$ & $X^2-Y^2$ & $XY$ & $XZ$ & $YZ$ \\ \hline
		\multirow{5}{*}{Cr$_1$} & $3Z^2-R^2$ & 37 & $-$14 & $-$9 & $-$77 & 2 \\
		& $X^2-Y^2$ & $-$14 & 21 & $-$3 & $-$134 & $-$9 \\
		& $XY$ & $-$9 & $-$3 & 14 & $-$16 & 78 \\
		& $XZ$ & $-$77 & $-$134 & $-$16 & $-$13 & $-$16 \\
		& $YZ$ & 2 & $-$9 & 78 & $-$16 & 14 \\  \hline 
		\multirow{5}{*}{Cr$_2$} & $3Z^2-R^2$ & 9 & 8 & 54 & $-$13 & $-$10 \\
		& $X^2-Y^2$ & $-$8 & 0 & $-$6 & 2 & 0 \\
		& $XY$ & 54 & 6 & 0 & 9 & 0 \\
		& $XZ$ & $-$10 & 0 & 0 & $-$6 & $-$26 \\
		& $YZ$ & $-$13 & $-$2 & 9 & $-$37 & $-$6 \\ \hline 
		\multirow{5}{*}{Cr$_3$} & $3Z^2-R^2$ & 21 & 21 & 8 & $-$2 & $-$7 \\
		& $X^2-Y^2$ & 21 & $-$3 & $-$3 & $-$8 & 13 \\
		& $XY$ & 8 & $-$3 & 8 & $-$10 & 9 \\
		& $XZ$ & $-$2 & $-$8 & $-$10 & 8 & 9 \\
		& $YZ$ & $-$7 & 13 & 9 & 9 & $-$29 \\ 
		\hline\hline
	\end{tabular}
	\label{tbhopping_Cr}
\end{table}

Fig.~\ref{CrI3_dos} confirms the Cr$^{3+}$ $S$ = 3/2 state with the formal $t_{2g}^3e_{g}^0$, and it shows a strong Cr $3d$-I $5p$ hybridization. In particular, the large I $5p$ component in the topmost valence bands and the major Cr $3d$ states in the conduction bands suggest that CrI$_3$ is a charge-transfer insulator, which is confirmed by the GGA + $U$ and the hybrid functional calculations both giving quite similar results, see Figs.~\ref{CrI3_dos}(b) and S1(c,d) for a comparison. 
We also carried out GGA + $U$ calculations for the four different magnetic structures (see Fig.~\ref{4Jstructure}) and find that the FM solution is most favorable and has a lower total energy than the other three different AF solutions by 9-17 meV/f.u. as shown in Table~\ref{tb3J}. Then, the three exchange parameters for CrI$_3$ monolayer are determined as $J_1 = 2.56$ meV, $J_2 = 0.39$ meV, and $J_3 = -0.04$ meV, which show the major 1NN FM coupling, the much weaker 2NN FM, and the negligible 3NN exchange.

\begin{figure}[t]
	\centering 
	\includegraphics[width=8.5cm]{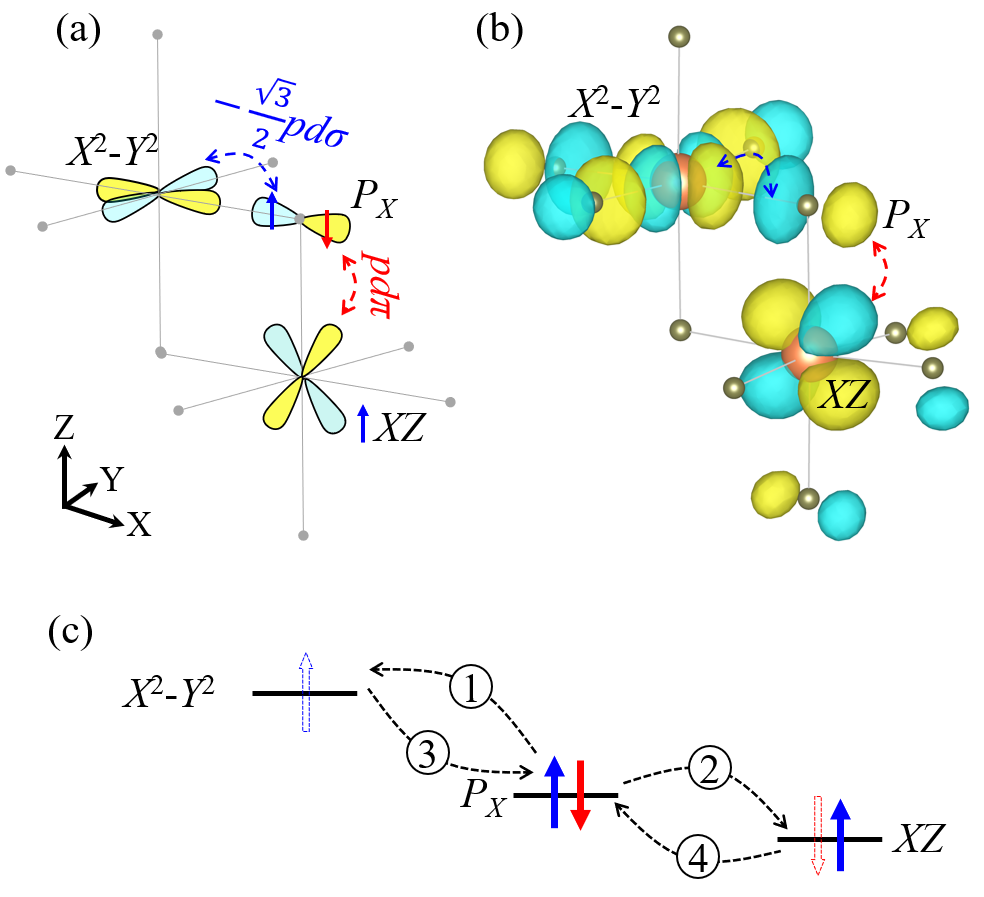}
	\centering
	\caption{(a) The hybridization between the I $p_X$ orbital and the $X^2-Y^2$ orbital, as well as between the I $p_X$ orbital and the $XZ$ orbital in CrI$_3$. (b) The corresponding Wannier orbitals. (c) The indirect ($X^2-Y^2$)-($p_X$)-($XZ$) hopping channels lead to FM coupling.}
	\label{CrI3_x2_xz_1}
\end{figure}

We now perform Wannier function analysis to identify the hopping channels in CrI$_3$, as seen in Fig. S11 in SM~\cite{SM}, to understand its FM order.
We first examine the primary hopping channels related to the 1NN FM coupling in the CrI$_3$ monolayer, which has a 1NN distance of 4.03~$\angstrom$ (compared to 3.42~$\angstrom$ in VPS$_3$). The larger 1NN Cr-Cr distance, combined with the smaller ionic radius of Cr$^{3+}$ (0.615~$\angstrom$) compared to V$^{2+}$ (0.79~$\angstrom$), weakens direct $d$-$d$ hopping in CrI$_3$. Then, indirect $d$-$p$-$d$ hopping may play a dominant role in determining the 1NN FM coupling. As shown in Table~\ref{tbhopping_Cr}, the diagonal hopping integral between two occupied $t_{2g}$ orbitals is only about 14 meV, being much smaller than the off-diagonal hopping integrals, such as 134 meV between the $XZ$ and $X^2-Y^2$. This implies that the predominant 1NN FM coupling in CrI$_3$ arises from the indirect $d$-$p$-$d$ superexchange rather than the direct $d$-$d$ exchange as in the aforementioned VPS$_3$. 

\begin{figure}[t]
	\centering 
	\includegraphics[width=8.5cm]{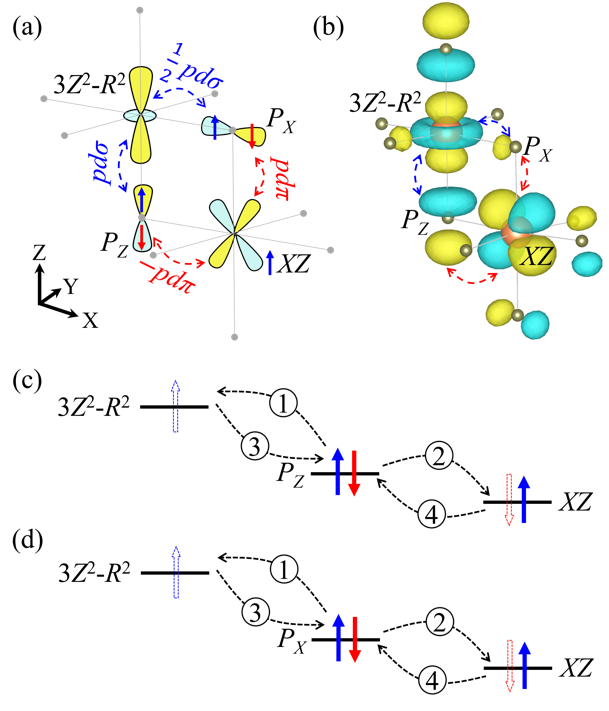}
	\centering
	\caption{(a) The hybridization between the I $p_X$ orbital and the $3Z^2-R^2$ orbital, between the I $p_X$ and the $XZ$, between the I $p_Z$ and the $3Z^2-R^2$, and between the I $p_Z$ and the $XZ$ in CrI$_3$. (b) The corresponding Wannier orbitals. (c) The indirect ($3Z^2-R^2$)-($p_Z$)-($XZ$) and ($3Z^2-R^2$)-($p_X$)-($XZ$) hopping channels both lead to FM coupling.}
	\label{CrI3_z2_xz_1}
\end{figure}

In the edge-sharing octahedra, as shown in Fig.~S12 in SM~\cite{SM}, the $XZ$ and $X^2-Y^2$ orbitals exhibit limited direct $d$-$d$ hybridization, with a contribution only of $-\frac{3}{8}dd\sigma + \frac{3}{8}dd\delta$. Given the larger 1NN Cr-Cr distance (4.03$\angstrom$) and the smaller ionic radius of Cr$^{3+}$ (0.615~$\angstrom$), this limited direct $d$-$d$ hybridization would be insignificant.
In contrast, the $XZ$ and $X^2-Y^2$ orbitals display strong $pd\pi$ and $-\frac{\sqrt{3}}{2}pd\sigma$ hybridizations with adjacent I $p_X$ orbitals, as seen in Fig.~\ref{CrI3_x2_xz_1}(a). Moreover, significant contributions from the I $p_X$ orbitals are observed in the $XZ$ and $X^2-Y^2$-like MLWFs, as shown in Fig.~\ref{CrI3_x2_xz_1}(b). Thus, the largest 1NN hopping integral between the $XZ$ and $X^2-Y^2$ orbitals in CrI$_3$ originates from the indirect $d$-$p$-$d$ hopping, with the effective hopping integral of ($-\frac{\sqrt{3}}{2}pd\sigma \cdot pd\pi$)/$\Delta$ ($\Delta$ being the charge transfer energy in CrI$_3$).
As shown in Fig.~\ref{CrI3_x2_xz_1}(c), the indirect ($X^2-Y^2$)-($p_X$)-($XZ$) superexchange channel contributes a lot to the 1NN FM coupling.

\begin{figure}[t]
	\centering 
	\includegraphics[width=8.5cm]{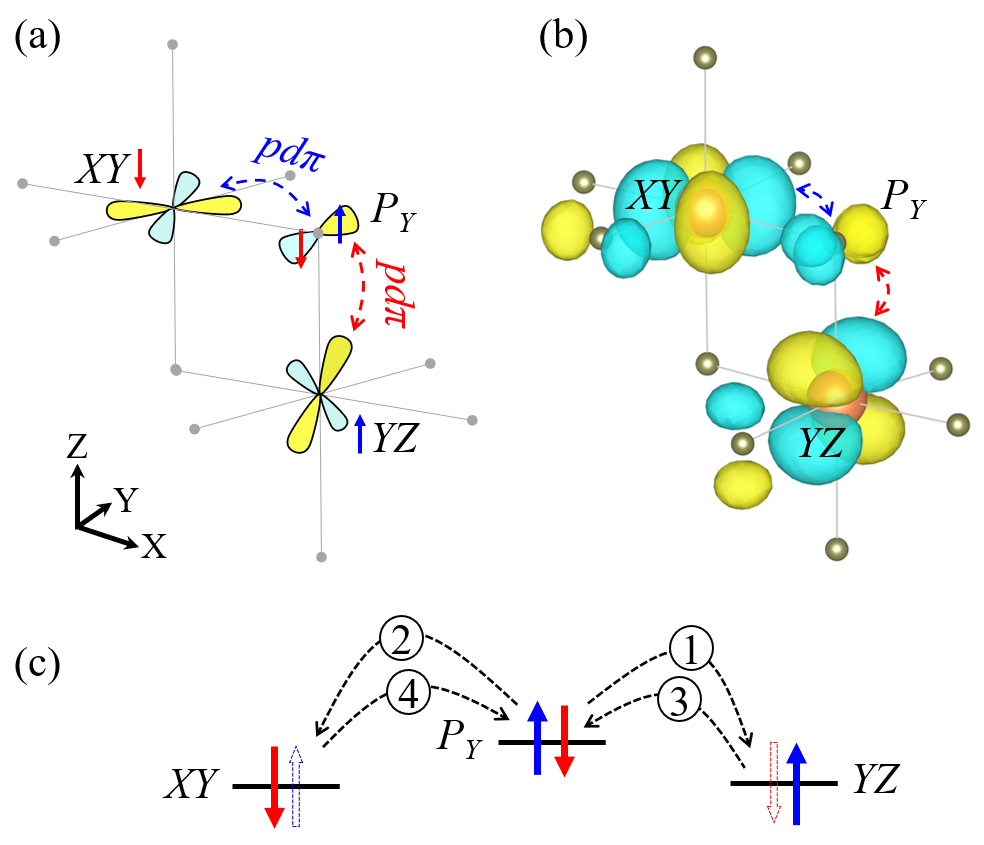}
	\centering
	\caption{(a) The hybridization between the I $p_Y$ orbital and the $XY$ orbital, and between the I $p_Y$ and the $YZ$. (b) The corresponding Wannier orbitals in CrI$_3$. (c) The indirect ($XY$)-($p_Y$)-($YZ$) hopping channels lead to AF coupling.}
	\label{CrI3_xy_yz_1}
\end{figure}

Next, we discuss the hopping integral of 77 meV between the $XZ$ and $3Z^2-R^2$ orbitals, which was previously treated as a key superexchange channel responsible for the FM coupling in CrI$_3$~\cite{Huang_2018_jacs,Kim2019,Xue2019}. As shown in Fig.~\ref{CrI3_z2_xz_1}(a), the $XZ$ and $3Z^2-R^2$ orbitals form $-pd\pi$ and $pd\sigma$ hybridizations with the adjacent I $p_Z$ orbital, respectively. Consequently, the hybridization through the ($3Z^2-R^2$)-($p_Z$)-($XZ$) channel, characterized by ($-pd\sigma \cdot pd\pi$)/$\Delta$, is larger in strength than the above ($X^2-Y^2$)-($p_X$)-($XZ$) channel with ($-\frac{\sqrt{3}}{2}pd\sigma \cdot pd\pi$)/$\Delta$.
However, the hopping integral between the $XZ$ and $3Z^2-R^2$ orbitals (77 meV) is smaller than that between the $XZ$ and $X^2-Y^2$ orbitals (134 meV). This discrepancy arises because the hopping between the $XZ$ and $3Z^2-R^2$ orbitals occurs not only through the adjacent I $p_Z$ orbital but also through the other adjacent I $p_X$ orbital. The $XZ$ and $3Z^2-R^2$ orbitals form $pd\pi$ and $\frac{1}{2}pd\sigma$ hybridizations with the other adjacent I $p_X$ orbitals. The hybridization through the ($3Z^2-R^2$)-($p_X$)-($XZ$) channel is characterized by ($\frac{1}{2}pd\sigma \cdot pd\pi$)/$\Delta$.
As a result, the total hopping between the $XZ$ and $3Z^2-R^2$ orbitals consists of these two channels: $\left(-pd\sigma \cdot pd\pi + \frac{1}{2}pd\sigma \cdot pd\pi\right)/ \Delta = (-\frac{1}{2}pd\sigma \cdot pd\pi)/ \Delta$. From this, we infer that the hopping integral between the $XZ$ and $3Z^2-R^2$ orbitals is about $\frac{1}{\sqrt{3}} \approx 0.577$ times that of the $XZ$ and $X^2-Y^2$ orbitals, which is in excellent agreement with our Wannier function results, i.e., $77/134 \approx 0.575$.
As shown in Figs.~\ref{CrI3_z2_xz_1}(c) and~\ref{CrI3_z2_xz_1}(d), considering virtual charge fluctuations, local Hund$'$s exchange, and the Pauli exclusion principle, both the ($3Z^2-R^2$)-($p_Z$)-($XZ$) and ($3Z^2-R^2$)-($p_X$)-($XZ$) superexchange channels would individually lead to FM coupling. However, as the combined effect of the ($3Z^2-R^2$)-($p_Z$, $p_X$)-($XZ$) channels should be seriously treated, their anti-phase contributions ultimately give rise to the partially reduced FM exchange. Thus, we find that the previously assumed key superexchange channel ($XZ$)-($p_Z$)-($3Z^2-R^2$) is not most important, but instead we propose the new and most important superexchange channel ($XZ$)-($p_X$)-($X^2-Y^2$) as depicted in Fig.~\ref{CrI3_x2_xz_1}.

Note, however, that besides the above FM superexchange channels between the occupied $t_{2g}$ and unoccupied $e_g$ orbitals, AF superexchange channels exist between two occupied $t_{2g}$ orbitals. For example, as shown in Table~\ref{tbhopping_Cr}, the hopping integral between the $XY$ and $YZ$ orbitals is 78 meV. Fig.~\ref{CrI3_xy_yz_1}(a) illustrates that both the $XY$ and $YZ$ orbitals form $pd\pi$ hybridizations with the adjacent I $p_Y$ orbitals. Consequently, the hybridization in the ($XY$)-($p_Y$)-($YZ$) channels is ($pd\pi \cdot pd\pi$)/$\Delta$, thereby giving weaker AF coupling.
As depicted in Fig.~\ref{CrI3_xy_yz_1}(c), when considering virtual charge fluctuations, local Hund$'$s exchange, and the Pauli exclusion principle, the ($XY$)-($p_Y$)-($YZ$) superexchange channel gives rise to AF coupling.
Thus, the 1NN FM coupling in CrI$_3$ primarily arises from the indirect FM superexchange involving the occupied $t_{2g}$ and unoccupied $e_g$ orbitals, which is partially reduced by the indirect AF superexchange between two occupied $t_{2g}$ orbitals.

As shown in Table~\ref{tbhopping_Cr}, the hopping parameters for the 2NN (6.99 $\angstrom$) are apparently smaller, and those for the 3NN (8.07 $\angstrom$) are even much smaller, both of which are in line with our DFT calculations of $J_2 = 0.39$ meV and $J_3 = -0.04$ meV being much smaller than $J_1$ = 2.56 meV. Note that unlike VPS$_3$ where the P-P dimer mediates the 3NN magnetic exchange (Fig. S9 in SM), CrI$_3$ has a void space at the center of the honeycomb Cr spin lattice (see Figs. 1 and 3(a)) and thus here the 3NN magnetic exchange is negligibly weak.

We also calculated the experimental out-of-plane MA of the CrI$_3$ monolayer. Although the shape MAE of the FM CrI$_3$ monolayer favors in-plane magnetization, the strong SOC effect from the I 5$p$ orbitals induces an out-of-plane exchange anisotropy, which ultimately determines the easy out-of-plane magnetization, see Fig. 7(b). This behavior was extensively discussed, e.g., in Refs.~\cite{Lado2017,Kim2019}.
Using Eq.~\ref{eq5}, which seems sufficient (without inclusion of the biquadratic exchange~\cite{Kartsev_2020}), together with the magnetic couplings $J_1 = 2.56$ meV, $J_2 = 0.39$ meV, and $J_3 = -0.04$ meV, and the calculated MA value of $D = 0.21$ meV, our PTMC simulations yield the $T_{\rm C}$ of 48 K for the CrI$_3$ monolayer, which is in good agreement with the experimental 46 K.

\section*{IV. Conclusion}

We have studied the electronic structure and in particular the contrasting magnetism of VPS$_3$ and CrI$_3$ monolayers both in the common honeycomb $S$=3/2 spin lattice. We find that VPS$_3$ is a Mott-Hubbard insulator and the direct 1NN AF exchange plays a major role in determining its N$\acute{e}$el AF behavior. The V vacancies against the less common V$^{2+}$ charge state suppress the otherwise strong AF exchange and then remarkably reduce the N$\acute{e}$el temperature in accord with the experimental observation. In contrast, CrI$_3$ is a charge-transfer insulator, and the indirect 1NN Cr $3d$-I $5p$-Cr $3d$ superexchange is dominant and consists of the FM superexchange via the occupied $t_{2g}$ orbitals and the unoccupied $e_g$ orbitals, and the relatively weak AF superexchange via the occupied $t_{2g}$ orbitals. Moreover, the AF VPS$_3$ has a weak shape anisotropy and the FM CrI$_3$ has a relatively strong exchange anisotropy, both favoring the out-of-plane magnetization. Using our calculated exchange parameters and the magnetic anisotropy energy, our PTMC simulations well reproduce the experimental $T_{\rm N}$ for the V deficient VPS$_3$ and the $T_{\rm C}$ for CrI$_3$. This work provides a comprehensive understanding of the two-dimensional magnetism and a new insight into the FM exchange of the extensively studied CrI$_3$, by a careful study of the different magnetic exchange channels.

\section*{Acknowledgements}
This work was supported by National Natural Science Foundation of China (Grants No. 12104307, No. 12174062, and No. 12241402), and by Innovation Program for Quantum Science and Technology (2024ZD0300102). 

\bibliography{VPS3_CrI3.bib}

\begin{thebibliography}{38}%
\makeatletter
\providecommand \@ifxundefined [1]{%
 \@ifx{#1\undefined}
}%
\providecommand \@ifnum [1]{%
 \ifnum #1\expandafter \@firstoftwo
 \else \expandafter \@secondoftwo
 \fi
}%
\providecommand \@ifx [1]{%
 \ifx #1\expandafter \@firstoftwo
 \else \expandafter \@secondoftwo
 \fi
}%
\providecommand \natexlab [1]{#1}%
\providecommand \enquote  [1]{``#1''}%
\providecommand \bibnamefont  [1]{#1}%
\providecommand \bibfnamefont [1]{#1}%
\providecommand \citenamefont [1]{#1}%
\providecommand \href@noop [0]{\@secondoftwo}%
\providecommand \href [0]{\begingroup \@sanitize@url \@href}%
\providecommand \@href[1]{\@@startlink{#1}\@@href}%
\providecommand \@@href[1]{\endgroup#1\@@endlink}%
\providecommand \@sanitize@url [0]{\catcode `\\12\catcode `\$12\catcode
  `\&12\catcode `\#12\catcode `\^12\catcode `\_12\catcode `\%12\relax}%
\providecommand \@@startlink[1]{}%
\providecommand \@@endlink[0]{}%
\providecommand \url  [0]{\begingroup\@sanitize@url \@url }%
\providecommand \@url [1]{\endgroup\@href {#1}{\urlprefix }}%
\providecommand \urlprefix  [0]{URL }%
\providecommand \Eprint [0]{\href }%
\providecommand \doibase [0]{https://doi.org/}%
\providecommand \selectlanguage [0]{\@gobble}%
\providecommand \bibinfo  [0]{\@secondoftwo}%
\providecommand \bibfield  [0]{\@secondoftwo}%
\providecommand \translation [1]{[#1]}%
\providecommand \BibitemOpen [0]{}%
\providecommand \bibitemStop [0]{}%
\providecommand \bibitemNoStop [0]{.\EOS\space}%
\providecommand \EOS [0]{\spacefactor3000\relax}%
\providecommand \BibitemShut  [1]{\csname bibitem#1\endcsname}%
\let\auto@bib@innerbib\@empty
\bibitem [{\citenamefont {Huang}\ \emph {et~al.}(2017)\citenamefont {Huang},
  \citenamefont {Clark}, \citenamefont {Navarro-Moratalla}, \citenamefont
  {Klein}, \citenamefont {Cheng}, \citenamefont {Seyler}, \citenamefont
  {Zhong}, \citenamefont {Schmidgall}, \citenamefont {McGuire}, \citenamefont
  {Cobden}, \citenamefont {Yao}, \citenamefont {Xiao}, \citenamefont
  {Jarillo-Herrero},\ and\ \citenamefont {Xu}}]{Huang2017}%
  \BibitemOpen
  \bibfield  {author} {\bibinfo {author} {\bibfnamefont {B.}~\bibnamefont
  {Huang}}, \bibinfo {author} {\bibfnamefont {G.}~\bibnamefont {Clark}},
  \bibinfo {author} {\bibfnamefont {E.}~\bibnamefont {Navarro-Moratalla}},
  \bibinfo {author} {\bibfnamefont {D.~R.}\ \bibnamefont {Klein}}, \bibinfo
  {author} {\bibfnamefont {R.}~\bibnamefont {Cheng}}, \bibinfo {author}
  {\bibfnamefont {K.~L.}\ \bibnamefont {Seyler}}, \bibinfo {author}
  {\bibfnamefont {D.}~\bibnamefont {Zhong}}, \bibinfo {author} {\bibfnamefont
  {E.}~\bibnamefont {Schmidgall}}, \bibinfo {author} {\bibfnamefont {M.~A.}\
  \bibnamefont {McGuire}}, \bibinfo {author} {\bibfnamefont {D.~H.}\
  \bibnamefont {Cobden}}, \bibinfo {author} {\bibfnamefont {W.}~\bibnamefont
  {Yao}}, \bibinfo {author} {\bibfnamefont {D.}~\bibnamefont {Xiao}}, \bibinfo
  {author} {\bibfnamefont {P.}~\bibnamefont {Jarillo-Herrero}},\ and\ \bibinfo
  {author} {\bibfnamefont {X.}~\bibnamefont {Xu}},\ }\bibfield  {title}
  {\bibinfo {title} {Layer-dependent ferromagnetism in a van der {Waals}
  crystal down to the monolayer limit},\ }\href
  {https://doi.org/10.1038/nature22391} {\bibfield  {journal} {\bibinfo
  {journal} {Nature}\ }\textbf {\bibinfo {volume} {546}},\ \bibinfo {pages}
  {270} (\bibinfo {year} {2017})}\BibitemShut {NoStop}%
\bibitem [{\citenamefont {Gong}\ \emph {et~al.}(2017)\citenamefont {Gong},
  \citenamefont {Li}, \citenamefont {Li}, \citenamefont {Ji}, \citenamefont
  {Stern}, \citenamefont {Xia}, \citenamefont {Cao}, \citenamefont {Bao},
  \citenamefont {Wang}, \citenamefont {Wang}, \citenamefont {Qiu},
  \citenamefont {Cava}, \citenamefont {Louie}, \citenamefont {Xia},\ and\
  \citenamefont {Zhang}}]{Gong2017}%
  \BibitemOpen
  \bibfield  {author} {\bibinfo {author} {\bibfnamefont {C.}~\bibnamefont
  {Gong}}, \bibinfo {author} {\bibfnamefont {L.}~\bibnamefont {Li}}, \bibinfo
  {author} {\bibfnamefont {Z.}~\bibnamefont {Li}}, \bibinfo {author}
  {\bibfnamefont {H.}~\bibnamefont {Ji}}, \bibinfo {author} {\bibfnamefont
  {A.}~\bibnamefont {Stern}}, \bibinfo {author} {\bibfnamefont
  {Y.}~\bibnamefont {Xia}}, \bibinfo {author} {\bibfnamefont {T.}~\bibnamefont
  {Cao}}, \bibinfo {author} {\bibfnamefont {W.}~\bibnamefont {Bao}}, \bibinfo
  {author} {\bibfnamefont {C.}~\bibnamefont {Wang}}, \bibinfo {author}
  {\bibfnamefont {Y.}~\bibnamefont {Wang}}, \bibinfo {author} {\bibfnamefont
  {Z.~Q.}\ \bibnamefont {Qiu}}, \bibinfo {author} {\bibfnamefont {R.~J.}\
  \bibnamefont {Cava}}, \bibinfo {author} {\bibfnamefont {S.~G.}\ \bibnamefont
  {Louie}}, \bibinfo {author} {\bibfnamefont {J.}~\bibnamefont {Xia}},\ and\
  \bibinfo {author} {\bibfnamefont {X.}~\bibnamefont {Zhang}},\ }\bibfield
  {title} {\bibinfo {title} {Discovery of intrinsic ferromagnetism in
  two-dimensional van der {Waals} crystals},\ }\href
  {https://www.nature.com/articles/nature22060} {\bibfield  {journal} {\bibinfo
   {journal} {Nature}\ }\textbf {\bibinfo {volume} {546}},\ \bibinfo {pages}
  {265} (\bibinfo {year} {2017})}\BibitemShut {NoStop}%
\bibitem [{\citenamefont {Mermin}\ and\ \citenamefont
  {Wagner}(1966)}]{Mermin1966}%
  \BibitemOpen
  \bibfield  {author} {\bibinfo {author} {\bibfnamefont {N.~D.}\ \bibnamefont
  {Mermin}}\ and\ \bibinfo {author} {\bibfnamefont {H.}~\bibnamefont
  {Wagner}},\ }\bibfield  {title} {\bibinfo {title} {Absence of ferromagnetism
  or antiferromagnetism in one- or two-dimensional isotropic {Heisenberg}
  models},\ }\href {https://link.aps.org/doi/10.1103/PhysRevLett.17.1133}
  {\bibfield  {journal} {\bibinfo  {journal} {Phys. Rev. Lett.}\ }\textbf
  {\bibinfo {volume} {17}},\ \bibinfo {pages} {1133} (\bibinfo {year}
  {1966})}\BibitemShut {NoStop}%
\bibitem [{\citenamefont {Lado}\ and\ \citenamefont
  {Fern{\'a}ndez-Rossier}(2017)}]{Lado2017}%
  \BibitemOpen
  \bibfield  {author} {\bibinfo {author} {\bibfnamefont {J.~L.}\ \bibnamefont
  {Lado}}\ and\ \bibinfo {author} {\bibfnamefont {J.}~\bibnamefont
  {Fern{\'a}ndez-Rossier}},\ }\bibfield  {title} {\bibinfo {title} {On the
  origin of magnetic anisotropy in two dimensional {CrI}$_{3}$},\ }\href
  {https://iopscience.iop.org/article/10.1088/2053-1583/aa75ed} {\bibfield
  {journal} {\bibinfo  {journal} {2D Mater.}\ }\textbf {\bibinfo {volume}
  {4}},\ \bibinfo {pages} {035002} (\bibinfo {year} {2017})}\BibitemShut
  {NoStop}%
\bibitem [{\citenamefont {Kim}\ \emph {et~al.}(2019)\citenamefont {Kim},
  \citenamefont {Kim}, \citenamefont {Ko}, \citenamefont {Seo}, \citenamefont
  {Kim}, \citenamefont {Jang}, \citenamefont {Kim}, \citenamefont {Kim},
  \citenamefont {Cheong},\ and\ \citenamefont {Park}}]{Kim2019}%
  \BibitemOpen
  \bibfield  {author} {\bibinfo {author} {\bibfnamefont {D.-H.}\ \bibnamefont
  {Kim}}, \bibinfo {author} {\bibfnamefont {K.}~\bibnamefont {Kim}}, \bibinfo
  {author} {\bibfnamefont {K.-T.}\ \bibnamefont {Ko}}, \bibinfo {author}
  {\bibfnamefont {J.}~\bibnamefont {Seo}}, \bibinfo {author} {\bibfnamefont
  {J.~S.}\ \bibnamefont {Kim}}, \bibinfo {author} {\bibfnamefont {T.-H.}\
  \bibnamefont {Jang}}, \bibinfo {author} {\bibfnamefont {Y.}~\bibnamefont
  {Kim}}, \bibinfo {author} {\bibfnamefont {J.-Y.}\ \bibnamefont {Kim}},
  \bibinfo {author} {\bibfnamefont {S.-W.}\ \bibnamefont {Cheong}},\ and\
  \bibinfo {author} {\bibfnamefont {J.-H.}\ \bibnamefont {Park}},\ }\bibfield
  {title} {\bibinfo {title} {Giant magnetic anisotropy induced by ligand ${LS}$
  coupling in layered {Cr} compounds},\ }\href
  {https://link.aps.org/doi/10.1103/PhysRevLett.122.207201} {\bibfield
  {journal} {\bibinfo  {journal} {Phys. Rev. Lett.}\ }\textbf {\bibinfo
  {volume} {122}},\ \bibinfo {pages} {207201} (\bibinfo {year}
  {2019})}\BibitemShut {NoStop}%
\bibitem [{\citenamefont {Yang}\ \emph {et~al.}(2020)\citenamefont {Yang},
  \citenamefont {Fan}, \citenamefont {Wang}, \citenamefont {Khomskii},\ and\
  \citenamefont {Wu}}]{Yang2020}%
  \BibitemOpen
  \bibfield  {author} {\bibinfo {author} {\bibfnamefont {K.}~\bibnamefont
  {Yang}}, \bibinfo {author} {\bibfnamefont {F.}~\bibnamefont {Fan}}, \bibinfo
  {author} {\bibfnamefont {H.}~\bibnamefont {Wang}}, \bibinfo {author}
  {\bibfnamefont {D.~I.}\ \bibnamefont {Khomskii}},\ and\ \bibinfo {author}
  {\bibfnamefont {H.}~\bibnamefont {Wu}},\ }\bibfield  {title} {\bibinfo
  {title} {{VI}$_{3}$: A two-dimensional {Ising} ferromagnet},\ }\href
  {https://link.aps.org/doi/10.1103/PhysRevB.101.100402} {\bibfield  {journal}
  {\bibinfo  {journal} {Phys. Rev. B}\ }\textbf {\bibinfo {volume} {101}},\
  \bibinfo {pages} {100402(R)} (\bibinfo {year} {2020})}\BibitemShut {NoStop}%
\bibitem [{\citenamefont {Zhao}\ \emph {et~al.}(2021)\citenamefont {Zhao},
  \citenamefont {Liu}, \citenamefont {Hu}, \citenamefont {Jia}, \citenamefont
  {Cui}, \citenamefont {Wu}, \citenamefont {Whangbo},\ and\ \citenamefont
  {Ren}}]{Zhao2021}%
  \BibitemOpen
  \bibfield  {author} {\bibinfo {author} {\bibfnamefont {G.-D.}\ \bibnamefont
  {Zhao}}, \bibinfo {author} {\bibfnamefont {X.}~\bibnamefont {Liu}}, \bibinfo
  {author} {\bibfnamefont {T.}~\bibnamefont {Hu}}, \bibinfo {author}
  {\bibfnamefont {F.}~\bibnamefont {Jia}}, \bibinfo {author} {\bibfnamefont
  {Y.}~\bibnamefont {Cui}}, \bibinfo {author} {\bibfnamefont {W.}~\bibnamefont
  {Wu}}, \bibinfo {author} {\bibfnamefont {M.-H.}\ \bibnamefont {Whangbo}},\
  and\ \bibinfo {author} {\bibfnamefont {W.}~\bibnamefont {Ren}},\ }\bibfield
  {title} {\bibinfo {title} {Difference in magnetic anisotropy of the
  ferromagnetic monolayers {VI}$_{3}$ and {CrI}$_{3}$},\ }\href
  {https://link.aps.org/doi/10.1103/PhysRevB.103.014438} {\bibfield  {journal}
  {\bibinfo  {journal} {Phys. Rev. B}\ }\textbf {\bibinfo {volume} {103}},\
  \bibinfo {pages} {014438} (\bibinfo {year} {2021})}\BibitemShut {NoStop}%
\bibitem [{\citenamefont {Ni}\ \emph {et~al.}(2021)\citenamefont {Ni},
  \citenamefont {Haglund}, \citenamefont {Wang}, \citenamefont {Xu},
  \citenamefont {Bernhard}, \citenamefont {Mandrus}, \citenamefont {Qian},
  \citenamefont {Mele}, \citenamefont {Kane},\ and\ \citenamefont
  {Wu}}]{Ni2021}%
  \BibitemOpen
  \bibfield  {author} {\bibinfo {author} {\bibfnamefont {Z.}~\bibnamefont
  {Ni}}, \bibinfo {author} {\bibfnamefont {A.~V.}\ \bibnamefont {Haglund}},
  \bibinfo {author} {\bibfnamefont {H.}~\bibnamefont {Wang}}, \bibinfo {author}
  {\bibfnamefont {B.}~\bibnamefont {Xu}}, \bibinfo {author} {\bibfnamefont
  {C.}~\bibnamefont {Bernhard}}, \bibinfo {author} {\bibfnamefont {D.~G.}\
  \bibnamefont {Mandrus}}, \bibinfo {author} {\bibfnamefont {X.}~\bibnamefont
  {Qian}}, \bibinfo {author} {\bibfnamefont {E.~J.}\ \bibnamefont {Mele}},
  \bibinfo {author} {\bibfnamefont {C.~L.}\ \bibnamefont {Kane}},\ and\
  \bibinfo {author} {\bibfnamefont {L.}~\bibnamefont {Wu}},\ }\bibfield
  {title} {\bibinfo {title} {Imaging the {N$\acute{e}$el} {AF} vector switching
  in the monolayer antiferromagnet {MnPSe}$_{3}$ with strain-controlled {Ising}
  order},\ }\href {https://www.nature.com/articles/s41565-021-00885-5}
  {\bibfield  {journal} {\bibinfo  {journal} {Nat. Nanotechnol.}\ }\textbf
  {\bibinfo {volume} {16}},\ \bibinfo {pages} {782} (\bibinfo {year}
  {2021})}\BibitemShut {NoStop}%
\bibitem [{\citenamefont {Chu}\ \emph {et~al.}(2020)\citenamefont {Chu},
  \citenamefont {Roh}, \citenamefont {Island}, \citenamefont {Li},
  \citenamefont {Lee}, \citenamefont {Chen}, \citenamefont {Park},
  \citenamefont {Young}, \citenamefont {Lee},\ and\ \citenamefont
  {Hsieh}}]{Chu_2020}%
  \BibitemOpen
  \bibfield  {author} {\bibinfo {author} {\bibfnamefont {H.}~\bibnamefont
  {Chu}}, \bibinfo {author} {\bibfnamefont {C.~J.}\ \bibnamefont {Roh}},
  \bibinfo {author} {\bibfnamefont {J.~O.}\ \bibnamefont {Island}}, \bibinfo
  {author} {\bibfnamefont {C.}~\bibnamefont {Li}}, \bibinfo {author}
  {\bibfnamefont {S.}~\bibnamefont {Lee}}, \bibinfo {author} {\bibfnamefont
  {J.}~\bibnamefont {Chen}}, \bibinfo {author} {\bibfnamefont {J.-G.}\
  \bibnamefont {Park}}, \bibinfo {author} {\bibfnamefont {A.~F.}\ \bibnamefont
  {Young}}, \bibinfo {author} {\bibfnamefont {J.~S.}\ \bibnamefont {Lee}},\
  and\ \bibinfo {author} {\bibfnamefont {D.}~\bibnamefont {Hsieh}},\ }\bibfield
   {title} {\bibinfo {title} {Linear magnetoelectric phase in ultrathin
  {MnPS}$_{3}$ probed by optical second harmonic generation},\ }\href
  {https://link.aps.org/doi/10.1103/PhysRevLett.124.027601} {\bibfield
  {journal} {\bibinfo  {journal} {Phys. Rev. Lett.}\ }\textbf {\bibinfo
  {volume} {124}},\ \bibinfo {pages} {027601} (\bibinfo {year}
  {2020})}\BibitemShut {NoStop}%
\bibitem [{\citenamefont {Lee}\ \emph {et~al.}(2016)\citenamefont {Lee},
  \citenamefont {Lee}, \citenamefont {Ryoo}, \citenamefont {Kang},
  \citenamefont {Kim}, \citenamefont {Kim}, \citenamefont {Park}, \citenamefont
  {Park},\ and\ \citenamefont {Cheong}}]{Lee2016}%
  \BibitemOpen
  \bibfield  {author} {\bibinfo {author} {\bibfnamefont {J.-U.}\ \bibnamefont
  {Lee}}, \bibinfo {author} {\bibfnamefont {S.}~\bibnamefont {Lee}}, \bibinfo
  {author} {\bibfnamefont {J.~H.}\ \bibnamefont {Ryoo}}, \bibinfo {author}
  {\bibfnamefont {S.}~\bibnamefont {Kang}}, \bibinfo {author} {\bibfnamefont
  {T.~Y.}\ \bibnamefont {Kim}}, \bibinfo {author} {\bibfnamefont
  {P.}~\bibnamefont {Kim}}, \bibinfo {author} {\bibfnamefont {C.-H.}\
  \bibnamefont {Park}}, \bibinfo {author} {\bibfnamefont {J.-G.}\ \bibnamefont
  {Park}},\ and\ \bibinfo {author} {\bibfnamefont {H.}~\bibnamefont {Cheong}},\
  }\bibfield  {title} {\bibinfo {title} {Ising-type magnetic ordering in
  atomically thin {FePS}$_{3}$},\ }\href
  {https://pubs.acs.org/doi/10.1021/acs.nanolett.6b03052} {\bibfield  {journal}
  {\bibinfo  {journal} {Nano Lett.}\ }\textbf {\bibinfo {volume} {16}},\
  \bibinfo {pages} {7433} (\bibinfo {year} {2016})}\BibitemShut {NoStop}%
\bibitem [{\citenamefont {Kang}\ \emph {et~al.}(2020)\citenamefont {Kang},
  \citenamefont {Kim}, \citenamefont {Kim}, \citenamefont {Kim}, \citenamefont
  {Sim}, \citenamefont {Lee}, \citenamefont {Lee}, \citenamefont {Park},
  \citenamefont {Yun}, \citenamefont {Kim}, \citenamefont {Nag}, \citenamefont
  {Walters}, \citenamefont {Garcia-Fernandez}, \citenamefont {Li},
  \citenamefont {Chapon}, \citenamefont {Zhou}, \citenamefont {Son},
  \citenamefont {Kim}, \citenamefont {Cheong},\ and\ \citenamefont
  {Park}}]{Kang2020}%
  \BibitemOpen
  \bibfield  {author} {\bibinfo {author} {\bibfnamefont {S.}~\bibnamefont
  {Kang}}, \bibinfo {author} {\bibfnamefont {K.}~\bibnamefont {Kim}}, \bibinfo
  {author} {\bibfnamefont {B.~H.}\ \bibnamefont {Kim}}, \bibinfo {author}
  {\bibfnamefont {J.}~\bibnamefont {Kim}}, \bibinfo {author} {\bibfnamefont
  {K.~I.}\ \bibnamefont {Sim}}, \bibinfo {author} {\bibfnamefont {J.-U.}\
  \bibnamefont {Lee}}, \bibinfo {author} {\bibfnamefont {S.}~\bibnamefont
  {Lee}}, \bibinfo {author} {\bibfnamefont {K.}~\bibnamefont {Park}}, \bibinfo
  {author} {\bibfnamefont {S.}~\bibnamefont {Yun}}, \bibinfo {author}
  {\bibfnamefont {T.}~\bibnamefont {Kim}}, \bibinfo {author} {\bibfnamefont
  {A.}~\bibnamefont {Nag}}, \bibinfo {author} {\bibfnamefont {A.}~\bibnamefont
  {Walters}}, \bibinfo {author} {\bibfnamefont {M.}~\bibnamefont
  {Garcia-Fernandez}}, \bibinfo {author} {\bibfnamefont {J.}~\bibnamefont
  {Li}}, \bibinfo {author} {\bibfnamefont {L.}~\bibnamefont {Chapon}}, \bibinfo
  {author} {\bibfnamefont {K.-J.}\ \bibnamefont {Zhou}}, \bibinfo {author}
  {\bibfnamefont {Y.-W.}\ \bibnamefont {Son}}, \bibinfo {author} {\bibfnamefont
  {J.~H.}\ \bibnamefont {Kim}}, \bibinfo {author} {\bibfnamefont
  {H.}~\bibnamefont {Cheong}},\ and\ \bibinfo {author} {\bibfnamefont {J.-G.}\
  \bibnamefont {Park}},\ }\bibfield  {title} {\bibinfo {title} {Coherent
  many-body exciton in van der {Waals} antiferromagnet {NiPS}$_{3}$},\ }\href
  {https://doi.org/10.1038/s41586-020-2520-5} {\bibfield  {journal} {\bibinfo
  {journal} {Nature}\ }\textbf {\bibinfo {volume} {583}},\ \bibinfo {pages}
  {785} (\bibinfo {year} {2020})}\BibitemShut {NoStop}%
\bibitem [{\citenamefont {Ju}\ \emph {et~al.}(2021)\citenamefont {Ju},
  \citenamefont {Lee}, \citenamefont {Kim}, \citenamefont {Choi}, \citenamefont
  {Roh}, \citenamefont {Son}, \citenamefont {Park}, \citenamefont {Kim},
  \citenamefont {Jung}, \citenamefont {Kim}, \citenamefont {Kim}, \citenamefont
  {Park},\ and\ \citenamefont {Lee}}]{Ju2021}%
  \BibitemOpen
  \bibfield  {author} {\bibinfo {author} {\bibfnamefont {H.}~\bibnamefont
  {Ju}}, \bibinfo {author} {\bibfnamefont {Y.}~\bibnamefont {Lee}}, \bibinfo
  {author} {\bibfnamefont {K.-T.}\ \bibnamefont {Kim}}, \bibinfo {author}
  {\bibfnamefont {I.~H.}\ \bibnamefont {Choi}}, \bibinfo {author}
  {\bibfnamefont {C.~J.}\ \bibnamefont {Roh}}, \bibinfo {author} {\bibfnamefont
  {S.}~\bibnamefont {Son}}, \bibinfo {author} {\bibfnamefont {P.}~\bibnamefont
  {Park}}, \bibinfo {author} {\bibfnamefont {J.~H.}\ \bibnamefont {Kim}},
  \bibinfo {author} {\bibfnamefont {T.~S.}\ \bibnamefont {Jung}}, \bibinfo
  {author} {\bibfnamefont {J.~H.}\ \bibnamefont {Kim}}, \bibinfo {author}
  {\bibfnamefont {K.~H.}\ \bibnamefont {Kim}}, \bibinfo {author} {\bibfnamefont
  {J.-G.}\ \bibnamefont {Park}},\ and\ \bibinfo {author} {\bibfnamefont
  {J.~S.}\ \bibnamefont {Lee}},\ }\bibfield  {title} {\bibinfo {title}
  {Possible persistence of multiferroic order down to bilayer limit of van der
  waals material {NiI}$_2$},\ }\href
  {https://pubs.acs.org/doi/full/10.1021/acs.nanolett.1c01095} {\bibfield
  {journal} {\bibinfo  {journal} {Nano Lett.}\ }\textbf {\bibinfo {volume}
  {21}},\ \bibinfo {pages} {5126} (\bibinfo {year} {2021})}\BibitemShut
  {NoStop}%
\bibitem [{\citenamefont {Kim}\ \emph {et~al.}(2023)\citenamefont {Kim},
  \citenamefont {Kim}, \citenamefont {Park}, \citenamefont {Kim}, \citenamefont
  {Jeong}, \citenamefont {Ohira-Kawamura}, \citenamefont {Murai}, \citenamefont
  {Nakajima}, \citenamefont {Chernyshev}, \citenamefont {Mourigal},
  \citenamefont {Kim},\ and\ \citenamefont {Park}}]{kim2023}%
  \BibitemOpen
  \bibfield  {author} {\bibinfo {author} {\bibfnamefont {C.}~\bibnamefont
  {Kim}}, \bibinfo {author} {\bibfnamefont {S.}~\bibnamefont {Kim}}, \bibinfo
  {author} {\bibfnamefont {P.}~\bibnamefont {Park}}, \bibinfo {author}
  {\bibfnamefont {T.}~\bibnamefont {Kim}}, \bibinfo {author} {\bibfnamefont
  {J.}~\bibnamefont {Jeong}}, \bibinfo {author} {\bibfnamefont
  {S.}~\bibnamefont {Ohira-Kawamura}}, \bibinfo {author} {\bibfnamefont
  {N.}~\bibnamefont {Murai}}, \bibinfo {author} {\bibfnamefont
  {K.}~\bibnamefont {Nakajima}}, \bibinfo {author} {\bibfnamefont
  {A.}~\bibnamefont {Chernyshev}}, \bibinfo {author} {\bibfnamefont
  {M.}~\bibnamefont {Mourigal}}, \bibinfo {author} {\bibfnamefont {S.-J.}\
  \bibnamefont {Kim}},\ and\ \bibinfo {author} {\bibfnamefont {J.-G.}\
  \bibnamefont {Park}},\ }\bibfield  {title} {\bibinfo {title} {Bond-dependent
  anisotropy and magnon decay in cobalt-based kitaev triangular
  antiferromagnet},\ }\href
  {https://www.nature.com/articles/s41567-023-02180-7} {\bibfield  {journal}
  {\bibinfo  {journal} {Nat. Phys.}\ }\textbf {\bibinfo {volume} {19}},\
  \bibinfo {pages} {1624} (\bibinfo {year} {2023})}\BibitemShut {NoStop}%
\bibitem [{\citenamefont {Liu}\ \emph {et~al.}(2023)\citenamefont {Liu},
  \citenamefont {Li}, \citenamefont {Hu}, \citenamefont {Duan}, \citenamefont
  {Wang}, \citenamefont {Cai}, \citenamefont {Feng}, \citenamefont {Wang},
  \citenamefont {Liu}, \citenamefont {Hou}, \citenamefont {Liu}, \citenamefont
  {Zhang}, \citenamefont {Zhu}, \citenamefont {Niu}, \citenamefont {Zakharov},
  \citenamefont {Sheng},\ and\ \citenamefont {Yan}}]{Liu2023}%
  \BibitemOpen
  \bibfield  {author} {\bibinfo {author} {\bibfnamefont {C.}~\bibnamefont
  {Liu}}, \bibinfo {author} {\bibfnamefont {Z.}~\bibnamefont {Li}}, \bibinfo
  {author} {\bibfnamefont {J.}~\bibnamefont {Hu}}, \bibinfo {author}
  {\bibfnamefont {H.}~\bibnamefont {Duan}}, \bibinfo {author} {\bibfnamefont
  {C.}~\bibnamefont {Wang}}, \bibinfo {author} {\bibfnamefont {L.}~\bibnamefont
  {Cai}}, \bibinfo {author} {\bibfnamefont {S.}~\bibnamefont {Feng}}, \bibinfo
  {author} {\bibfnamefont {Y.}~\bibnamefont {Wang}}, \bibinfo {author}
  {\bibfnamefont {R.}~\bibnamefont {Liu}}, \bibinfo {author} {\bibfnamefont
  {D.}~\bibnamefont {Hou}}, \bibinfo {author} {\bibfnamefont {C.}~\bibnamefont
  {Liu}}, \bibinfo {author} {\bibfnamefont {R.}~\bibnamefont {Zhang}}, \bibinfo
  {author} {\bibfnamefont {L.}~\bibnamefont {Zhu}}, \bibinfo {author}
  {\bibfnamefont {Y.}~\bibnamefont {Niu}}, \bibinfo {author} {\bibfnamefont
  {A.~A.~A.}\ \bibnamefont {Zakharov}}, \bibinfo {author} {\bibfnamefont
  {Z.}~\bibnamefont {Sheng}},\ and\ \bibinfo {author} {\bibfnamefont
  {W.}~\bibnamefont {Yan}},\ }\bibfield  {title} {\bibinfo {title} {Probing the
  {N$\acute{e}$el}-type antiferromagnetic order and coherent magnon-exciton
  coupling in van der {Waals} {VPS}$_{3}$},\ }\href
  {https://doi.org/10.1002/adma.202300247} {\bibfield  {journal} {\bibinfo
  {journal} {Adv. Mater.}\ }\textbf {\bibinfo {volume} {35}},\ \bibinfo {pages}
  {2300247} (\bibinfo {year} {2023})}\BibitemShut {NoStop}%
\bibitem [{\citenamefont {Wildes}\ \emph {et~al.}(2017)\citenamefont {Wildes},
  \citenamefont {Simonet}, \citenamefont {Ressouche}, \citenamefont {Ballou},\
  and\ \citenamefont {McIntyre}}]{Wildes2017}%
  \BibitemOpen
  \bibfield  {author} {\bibinfo {author} {\bibfnamefont {A.~R.}\ \bibnamefont
  {Wildes}}, \bibinfo {author} {\bibfnamefont {V.}~\bibnamefont {Simonet}},
  \bibinfo {author} {\bibfnamefont {E.}~\bibnamefont {Ressouche}}, \bibinfo
  {author} {\bibfnamefont {R.}~\bibnamefont {Ballou}},\ and\ \bibinfo {author}
  {\bibfnamefont {G.~J.}\ \bibnamefont {McIntyre}},\ }\bibfield  {title}
  {\bibinfo {title} {The magnetic properties and structure of the
  quasi-two-dimensional antiferromagnet {CoPS}$_{3}$},\ }\href
  {https://iopscience.iop.org/article/10.1088/1361-648X/aa8a43} {\bibfield
  {journal} {\bibinfo  {journal} {J. Phys.: Condens. Matter}\ }\textbf
  {\bibinfo {volume} {29}},\ \bibinfo {pages} {455801} (\bibinfo {year}
  {2017})}\BibitemShut {NoStop}%
\bibitem [{\citenamefont {Bazazzadeh}\ \emph {et~al.}(2021)\citenamefont
  {Bazazzadeh}, \citenamefont {Hamdi}, \citenamefont {Haddadi}, \citenamefont
  {Khavasi}, \citenamefont {Sadeghi},\ and\ \citenamefont
  {Mohseni}}]{Bazazzadeh2021}%
  \BibitemOpen
  \bibfield  {author} {\bibinfo {author} {\bibfnamefont {N.}~\bibnamefont
  {Bazazzadeh}}, \bibinfo {author} {\bibfnamefont {M.}~\bibnamefont {Hamdi}},
  \bibinfo {author} {\bibfnamefont {F.}~\bibnamefont {Haddadi}}, \bibinfo
  {author} {\bibfnamefont {A.}~\bibnamefont {Khavasi}}, \bibinfo {author}
  {\bibfnamefont {A.}~\bibnamefont {Sadeghi}},\ and\ \bibinfo {author}
  {\bibfnamefont {S.~M.}\ \bibnamefont {Mohseni}},\ }\bibfield  {title}
  {\bibinfo {title} {Symmetry enhanced spin-{Nernst} effect in honeycomb
  antiferromagnetic transition metal trichalcogenide monolayers},\ }\href
  {https://link.aps.org/doi/10.1103/PhysRevB.103.014425} {\bibfield  {journal}
  {\bibinfo  {journal} {Phys. Rev. B}\ }\textbf {\bibinfo {volume} {103}},\
  \bibinfo {pages} {014425} (\bibinfo {year} {2021})}\BibitemShut {NoStop}%
\bibitem [{\citenamefont {Chittari}\ \emph {et~al.}(2016)\citenamefont
  {Chittari}, \citenamefont {Park}, \citenamefont {Lee}, \citenamefont {Han},
  \citenamefont {MacDonald}, \citenamefont {Hwang},\ and\ \citenamefont
  {Jung}}]{Chittari2016}%
  \BibitemOpen
  \bibfield  {author} {\bibinfo {author} {\bibfnamefont {B.~L.}\ \bibnamefont
  {Chittari}}, \bibinfo {author} {\bibfnamefont {Y.}~\bibnamefont {Park}},
  \bibinfo {author} {\bibfnamefont {D.}~\bibnamefont {Lee}}, \bibinfo {author}
  {\bibfnamefont {M.}~\bibnamefont {Han}}, \bibinfo {author} {\bibfnamefont
  {A.~H.}\ \bibnamefont {MacDonald}}, \bibinfo {author} {\bibfnamefont
  {E.}~\bibnamefont {Hwang}},\ and\ \bibinfo {author} {\bibfnamefont
  {J.}~\bibnamefont {Jung}},\ }\bibfield  {title} {\bibinfo {title} {Electronic
  and magnetic properties of single-layer ${M}${P}${X}_{3}$ metal phosphorous
  trichalcogenides},\ }\href
  {https://link.aps.org/doi/10.1103/PhysRevB.94.184428} {\bibfield  {journal}
  {\bibinfo  {journal} {Phys. Rev. B}\ }\textbf {\bibinfo {volume} {94}},\
  \bibinfo {pages} {184428} (\bibinfo {year} {2016})}\BibitemShut {NoStop}%
\bibitem [{\citenamefont {Kresse}\ and\ \citenamefont
  {Hafner}(1993)}]{Kresse1993}%
  \BibitemOpen
  \bibfield  {author} {\bibinfo {author} {\bibfnamefont {G.}~\bibnamefont
  {Kresse}}\ and\ \bibinfo {author} {\bibfnamefont {J.}~\bibnamefont
  {Hafner}},\ }\bibfield  {title} {\bibinfo {title} {{$Ab$} $initio$ molecular
  dynamics for liquid metals},\ }\href
  {https://link.aps.org/doi/10.1103/PhysRevB.47.558} {\bibfield  {journal}
  {\bibinfo  {journal} {Phys. Rev. B}\ }\textbf {\bibinfo {volume} {47}},\
  \bibinfo {pages} {558} (\bibinfo {year} {1993})}\BibitemShut {NoStop}%
\bibitem [{\citenamefont {Perdew}\ \emph {et~al.}(1996)\citenamefont {Perdew},
  \citenamefont {Burke},\ and\ \citenamefont {Ernzerhof}}]{Perdew1996}%
  \BibitemOpen
  \bibfield  {author} {\bibinfo {author} {\bibfnamefont {J.~P.}\ \bibnamefont
  {Perdew}}, \bibinfo {author} {\bibfnamefont {K.}~\bibnamefont {Burke}},\ and\
  \bibinfo {author} {\bibfnamefont {M.}~\bibnamefont {Ernzerhof}},\ }\bibfield
  {title} {\bibinfo {title} {Generalized gradient approximation made simple},\
  }\href {https://link.aps.org/doi/10.1103/PhysRevLett.77.3865} {\bibfield
  {journal} {\bibinfo  {journal} {Phys. Rev. Lett.}\ }\textbf {\bibinfo
  {volume} {77}},\ \bibinfo {pages} {3865} (\bibinfo {year}
  {1996})}\BibitemShut {NoStop}%
\bibitem [{\citenamefont {Klingen}\ \emph {et~al.}(1970)\citenamefont
  {Klingen}, \citenamefont {Eulenberger},\ and\ \citenamefont
  {Hahn}}]{Klingen1970}%
  \BibitemOpen
  \bibfield  {author} {\bibinfo {author} {\bibfnamefont {W.}~\bibnamefont
  {Klingen}}, \bibinfo {author} {\bibfnamefont {G.}~\bibnamefont
  {Eulenberger}},\ and\ \bibinfo {author} {\bibfnamefont {H.}~\bibnamefont
  {Hahn}},\ }\bibfield  {title} {\bibinfo {title} {Über
  hexachalkogeno-hypodiphosphate vom typ {M$_2$P$_2$X$_6$}},\ }\href
  {https://link.springer.com/article/10.1007/BF00590690} {\bibfield  {journal}
  {\bibinfo  {journal} {Naturwissenschaften}\ }\textbf {\bibinfo {volume}
  {57}},\ \bibinfo {pages} {88} (\bibinfo {year} {1970})}\BibitemShut {NoStop}%
\bibitem [{\citenamefont {McGuire}\ \emph {et~al.}(2015)\citenamefont
  {McGuire}, \citenamefont {Dixit}, \citenamefont {Cooper},\ and\ \citenamefont
  {Sales}}]{McGuire2015}%
  \BibitemOpen
  \bibfield  {author} {\bibinfo {author} {\bibfnamefont {M.~A.}\ \bibnamefont
  {McGuire}}, \bibinfo {author} {\bibfnamefont {H.}~\bibnamefont {Dixit}},
  \bibinfo {author} {\bibfnamefont {V.~R.}\ \bibnamefont {Cooper}},\ and\
  \bibinfo {author} {\bibfnamefont {B.~C.}\ \bibnamefont {Sales}},\ }\bibfield
  {title} {\bibinfo {title} {Coupling of crystal structure and magnetism in the
  layered, ferromagnetic insulator {CrI$_3$}},\ }\href
  {https://pubs.acs.org/doi/10.1021/cm504242t} {\bibfield  {journal} {\bibinfo
  {journal} {Chem. Mater.}\ }\textbf {\bibinfo {volume} {27}},\ \bibinfo
  {pages} {612} (\bibinfo {year} {2015})}\BibitemShut {NoStop}%
\bibitem [{SM()}]{SM}%
  \BibitemOpen
  \href@noop {} {}\bibinfo {note} {See Supplemental Material at
  http://link.aps.org/supplemental/*** for the convergence tests, the HSE06
  results for VPS$_3$ monolayer, the DFT calculated and Wannier interpolated
  band structures, Wannier functions and the hopping parameters for VPS$_3$ and
  CrI$_3$ monolayers, and the V vacancy results.}\BibitemShut {Stop}%
\bibitem [{\citenamefont {Anisimov}\ \emph {et~al.}(1997)\citenamefont
  {Anisimov}, \citenamefont {Aryasetiawan},\ and\ \citenamefont
  {Lichtenstein}}]{Anisimov1997}%
  \BibitemOpen
  \bibfield  {author} {\bibinfo {author} {\bibfnamefont {V.~I.}\ \bibnamefont
  {Anisimov}}, \bibinfo {author} {\bibfnamefont {F.}~\bibnamefont
  {Aryasetiawan}},\ and\ \bibinfo {author} {\bibfnamefont {A.~I.}\ \bibnamefont
  {Lichtenstein}},\ }\bibfield  {title} {\bibinfo {title} {First-principles
  calculations of the electronic structure and spectra of strongly correlated
  systems: the {LDA} + ${U}$ method},\ }\href
  {https://dx.doi.org/10.1088/0953-8984/9/4/002} {\bibfield  {journal}
  {\bibinfo  {journal} {J. Phys.: Condens. Matter}\ }\textbf {\bibinfo {volume}
  {9}},\ \bibinfo {pages} {767} (\bibinfo {year} {1997})}\BibitemShut {NoStop}%
\bibitem [{\citenamefont {Huang}\ \emph {et~al.}(2018)\citenamefont {Huang},
  \citenamefont {Feng}, \citenamefont {Wu}, \citenamefont {Ahmed},
  \citenamefont {Huang}, \citenamefont {Xiang}, \citenamefont {Deng},\ and\
  \citenamefont {Kan}}]{Huang_2018_jacs}%
  \BibitemOpen
  \bibfield  {author} {\bibinfo {author} {\bibfnamefont {C.}~\bibnamefont
  {Huang}}, \bibinfo {author} {\bibfnamefont {J.}~\bibnamefont {Feng}},
  \bibinfo {author} {\bibfnamefont {F.}~\bibnamefont {Wu}}, \bibinfo {author}
  {\bibfnamefont {D.}~\bibnamefont {Ahmed}}, \bibinfo {author} {\bibfnamefont
  {B.}~\bibnamefont {Huang}}, \bibinfo {author} {\bibfnamefont
  {H.}~\bibnamefont {Xiang}}, \bibinfo {author} {\bibfnamefont
  {K.}~\bibnamefont {Deng}},\ and\ \bibinfo {author} {\bibfnamefont
  {E.}~\bibnamefont {Kan}},\ }\bibfield  {title} {\bibinfo {title} {Toward
  intrinsic room-temperature ferromagnetism in two-dimensional
  semiconductors},\ }\href {https://pubs.acs.org/doi/10.1021/jacs.8b07879}
  {\bibfield  {journal} {\bibinfo  {journal} {J. Am. Chem. Soc.}\ }\textbf
  {\bibinfo {volume} {140}},\ \bibinfo {pages} {11519} (\bibinfo {year}
  {2018})}\BibitemShut {NoStop}%
\bibitem [{\citenamefont {Krukau}\ \emph {et~al.}(2006)\citenamefont {Krukau},
  \citenamefont {Vydrov}, \citenamefont {Izmaylov},\ and\ \citenamefont
  {Scuseria}}]{HSE06}%
  \BibitemOpen
  \bibfield  {author} {\bibinfo {author} {\bibfnamefont {A.~V.}\ \bibnamefont
  {Krukau}}, \bibinfo {author} {\bibfnamefont {O.~A.}\ \bibnamefont {Vydrov}},
  \bibinfo {author} {\bibfnamefont {A.~F.}\ \bibnamefont {Izmaylov}},\ and\
  \bibinfo {author} {\bibfnamefont {G.~E.}\ \bibnamefont {Scuseria}},\
  }\bibfield  {title} {\bibinfo {title} {Influence of the exchange screening
  parameter on the performance of screened hybrid functionals},\ }\href
  {https://doi.org/10.1063/1.2404663} {\bibfield  {journal} {\bibinfo
  {journal} {J. Chem. Phys.}\ }\textbf {\bibinfo {volume} {125}},\ \bibinfo
  {pages} {224106} (\bibinfo {year} {2006})}\BibitemShut {NoStop}%
\bibitem [{\citenamefont {Mostofi}\ \emph {et~al.}(2008)\citenamefont
  {Mostofi}, \citenamefont {Yates}, \citenamefont {Lee}, \citenamefont {Souza},
  \citenamefont {Vanderbilt},\ and\ \citenamefont {Marzari}}]{Mostofi2008}%
  \BibitemOpen
  \bibfield  {author} {\bibinfo {author} {\bibfnamefont {A.~A.}\ \bibnamefont
  {Mostofi}}, \bibinfo {author} {\bibfnamefont {J.~R.}\ \bibnamefont {Yates}},
  \bibinfo {author} {\bibfnamefont {Y.-S.}\ \bibnamefont {Lee}}, \bibinfo
  {author} {\bibfnamefont {I.}~\bibnamefont {Souza}}, \bibinfo {author}
  {\bibfnamefont {D.}~\bibnamefont {Vanderbilt}},\ and\ \bibinfo {author}
  {\bibfnamefont {N.}~\bibnamefont {Marzari}},\ }\bibfield  {title} {\bibinfo
  {title} {wannier90: A tool for obtaining maximally-localised {Wannier}
  functions},\ }\href
  {https://www.sciencedirect.com/science/article/pii/S0010465507004936}
  {\bibfield  {journal} {\bibinfo  {journal} {Comput. Phys. Commun.}\ }\textbf
  {\bibinfo {volume} {178}},\ \bibinfo {pages} {685} (\bibinfo {year}
  {2008})}\BibitemShut {NoStop}%
\bibitem [{\citenamefont {Marzari}\ \emph {et~al.}(2012)\citenamefont
  {Marzari}, \citenamefont {Mostofi}, \citenamefont {Yates}, \citenamefont
  {Souza},\ and\ \citenamefont {Vanderbilt}}]{Nicola2012}%
  \BibitemOpen
  \bibfield  {author} {\bibinfo {author} {\bibfnamefont {N.}~\bibnamefont
  {Marzari}}, \bibinfo {author} {\bibfnamefont {A.~A.}\ \bibnamefont
  {Mostofi}}, \bibinfo {author} {\bibfnamefont {J.~R.}\ \bibnamefont {Yates}},
  \bibinfo {author} {\bibfnamefont {I.}~\bibnamefont {Souza}},\ and\ \bibinfo
  {author} {\bibfnamefont {D.}~\bibnamefont {Vanderbilt}},\ }\bibfield  {title}
  {\bibinfo {title} {Maximally localized {Wannier} functions: Theory and
  applications},\ }\href {https://doi.org/10.1103/RevModPhys.84.1419}
  {\bibfield  {journal} {\bibinfo  {journal} {Rev. Mod. Phys.}\ }\textbf
  {\bibinfo {volume} {84}},\ \bibinfo {pages} {1419} (\bibinfo {year}
  {2012})}\BibitemShut {NoStop}%
\bibitem [{\citenamefont {Hukushima}\ and\ \citenamefont
  {Nemoto}(1996)}]{PTMC}%
  \BibitemOpen
  \bibfield  {author} {\bibinfo {author} {\bibfnamefont {K.}~\bibnamefont
  {Hukushima}}\ and\ \bibinfo {author} {\bibfnamefont {K.}~\bibnamefont
  {Nemoto}},\ }\bibfield  {title} {\bibinfo {title} {Exchange {Monte} {Carlo}
  method and application to spin glass simulations},\ }\href
  {https://doi.org/10.1143/JPSJ.65.1604} {\bibfield  {journal} {\bibinfo
  {journal} {J. Phys. Soc. Jpn.}\ }\textbf {\bibinfo {volume} {65}},\ \bibinfo
  {pages} {1604} (\bibinfo {year} {1996})}\BibitemShut {NoStop}%
\bibitem [{\citenamefont {Metropolis}\ and\ \citenamefont
  {Ulam}(1949)}]{Metropolis1949}%
  \BibitemOpen
  \bibfield  {author} {\bibinfo {author} {\bibfnamefont {N.}~\bibnamefont
  {Metropolis}}\ and\ \bibinfo {author} {\bibfnamefont {S.}~\bibnamefont
  {Ulam}},\ }\bibfield  {title} {\bibinfo {title} {The {Monte} {Carlo}
  method},\ }\href
  {https://www.tandfonline.com/doi/abs/10.1080/01621459.1949.10483310}
  {\bibfield  {journal} {\bibinfo  {journal} {J. Am. Stat. Assoc.}\ }\textbf
  {\bibinfo {volume} {44}},\ \bibinfo {pages} {335} (\bibinfo {year}
  {1949})}\BibitemShut {NoStop}%
\bibitem [{\citenamefont {Sivadas}\ \emph {et~al.}(2015)\citenamefont
  {Sivadas}, \citenamefont {Daniels}, \citenamefont {Swendsen}, \citenamefont
  {Okamoto},\ and\ \citenamefont {Xiao}}]{Sivadas_2015}%
  \BibitemOpen
  \bibfield  {author} {\bibinfo {author} {\bibfnamefont {N.}~\bibnamefont
  {Sivadas}}, \bibinfo {author} {\bibfnamefont {M.~W.}\ \bibnamefont
  {Daniels}}, \bibinfo {author} {\bibfnamefont {R.~H.}\ \bibnamefont
  {Swendsen}}, \bibinfo {author} {\bibfnamefont {S.}~\bibnamefont {Okamoto}},\
  and\ \bibinfo {author} {\bibfnamefont {D.}~\bibnamefont {Xiao}},\ }\bibfield
  {title} {\bibinfo {title} {Magnetic ground state of semiconducting
  transition-metal trichalcogenide monolayers},\ }\href
  {https://doi.org/10.1103/PhysRevB.91.235425} {\bibfield  {journal} {\bibinfo
  {journal} {Phys. Rev. B}\ }\textbf {\bibinfo {volume} {91}},\ \bibinfo
  {pages} {235425} (\bibinfo {year} {2015})}\BibitemShut {NoStop}%
\bibitem [{\citenamefont {Torelli}\ and\ \citenamefont
  {Olsen}(2019)}]{Torelli_2018}%
  \BibitemOpen
  \bibfield  {author} {\bibinfo {author} {\bibfnamefont {D.}~\bibnamefont
  {Torelli}}\ and\ \bibinfo {author} {\bibfnamefont {T.}~\bibnamefont
  {Olsen}},\ }\bibfield  {title} {\bibinfo {title} {Calculating critical
  temperatures for ferromagnetic order in two-dimensional materials},\ }\href
  {https://doi.org/10.1088/2053-1583/aaf06d} {\bibfield  {journal} {\bibinfo
  {journal} {2D Mater.}\ }\textbf {\bibinfo {volume} {6}},\ \bibinfo {pages}
  {015028} (\bibinfo {year} {2019})}\BibitemShut {NoStop}%
\bibitem [{\citenamefont {Yang}\ \emph {et~al.}(2024)\citenamefont {Yang},
  \citenamefont {Ning}, \citenamefont {Zhou}, \citenamefont {Lu}, \citenamefont
  {Ma}, \citenamefont {Liu}, \citenamefont {Pu},\ and\ \citenamefont
  {Wu}}]{Yang_2024}%
  \BibitemOpen
  \bibfield  {author} {\bibinfo {author} {\bibfnamefont {K.}~\bibnamefont
  {Yang}}, \bibinfo {author} {\bibfnamefont {Y.}~\bibnamefont {Ning}}, \bibinfo
  {author} {\bibfnamefont {Y.}~\bibnamefont {Zhou}}, \bibinfo {author}
  {\bibfnamefont {D.}~\bibnamefont {Lu}}, \bibinfo {author} {\bibfnamefont
  {Y.}~\bibnamefont {Ma}}, \bibinfo {author} {\bibfnamefont {L.}~\bibnamefont
  {Liu}}, \bibinfo {author} {\bibfnamefont {S.}~\bibnamefont {Pu}},\ and\
  \bibinfo {author} {\bibfnamefont {H.}~\bibnamefont {Wu}},\ }\bibfield
  {title} {\bibinfo {title} {Understanding the ising zigzag antiferromagnetism
  of {FePS}$_3$ and {FePSe}$_3$ monolayers},\ }\href
  {https://doi.org/10.1103/PhysRevB.110.024427} {\bibfield  {journal} {\bibinfo
   {journal} {Phys. Rev. B}\ }\textbf {\bibinfo {volume} {110}},\ \bibinfo
  {pages} {024427} (\bibinfo {year} {2024})}\BibitemShut {NoStop}%
\bibitem [{\citenamefont {Pandey}\ \emph {et~al.}(2021)\citenamefont {Pandey},
  \citenamefont {Kumar}, \citenamefont {Sarkar},\ and\ \citenamefont
  {Mahadevan}}]{Pandey_2021}%
  \BibitemOpen
  \bibfield  {author} {\bibinfo {author} {\bibfnamefont {S.~K.}\ \bibnamefont
  {Pandey}}, \bibinfo {author} {\bibfnamefont {A.}~\bibnamefont {Kumar}},
  \bibinfo {author} {\bibfnamefont {S.}~\bibnamefont {Sarkar}},\ and\ \bibinfo
  {author} {\bibfnamefont {P.}~\bibnamefont {Mahadevan}},\ }\bibfield  {title}
  {\bibinfo {title} {Understanding the ferromagnetic insulating state in
  {Cr}-doped {VO}$_{2}$ : Density functional and tight binding calculations},\
  }\href {https://doi.org/10.1103/PhysRevB.104.125110} {\bibfield  {journal}
  {\bibinfo  {journal} {Phys. Rev. B}\ }\textbf {\bibinfo {volume} {104}},\
  \bibinfo {pages} {125110} (\bibinfo {year} {2021})}\BibitemShut {NoStop}%
\bibitem [{\citenamefont {Coak}\ \emph
  {et~al.}(2019{\natexlab{a}})\citenamefont {Coak}, \citenamefont {Son},
  \citenamefont {Daisenberger}, \citenamefont {Hamidov}, \citenamefont
  {Haines}, \citenamefont {Alireza}, \citenamefont {Wildes}, \citenamefont
  {Liu}, \citenamefont {Saxena},\ and\ \citenamefont {Park}}]{Coak}%
  \BibitemOpen
  \bibfield  {author} {\bibinfo {author} {\bibfnamefont {M.~J.}\ \bibnamefont
  {Coak}}, \bibinfo {author} {\bibfnamefont {S.}~\bibnamefont {Son}}, \bibinfo
  {author} {\bibfnamefont {D.}~\bibnamefont {Daisenberger}}, \bibinfo {author}
  {\bibfnamefont {H.}~\bibnamefont {Hamidov}}, \bibinfo {author} {\bibfnamefont
  {C.~R.~S.}\ \bibnamefont {Haines}}, \bibinfo {author} {\bibfnamefont {P.~L.}\
  \bibnamefont {Alireza}}, \bibinfo {author} {\bibfnamefont {A.~R.}\
  \bibnamefont {Wildes}}, \bibinfo {author} {\bibfnamefont {C.}~\bibnamefont
  {Liu}}, \bibinfo {author} {\bibfnamefont {S.~S.}\ \bibnamefont {Saxena}},\
  and\ \bibinfo {author} {\bibfnamefont {J.-G.}\ \bibnamefont {Park}},\
  }\bibfield  {title} {\bibinfo {title} {Isostructural {Mott} transition in
  {2D} honeycomb antiferromagnet {V}$_{0.9}${PS}$_{3}$},\ }\href
  {https://doi.org/10.1038/s41535-019-0178-8} {\bibfield  {journal} {\bibinfo
  {journal} {npj Quantum Mater.}\ }\textbf {\bibinfo {volume} {4}},\ \bibinfo
  {pages} {38} (\bibinfo {year} {2019}{\natexlab{a}})}\BibitemShut {NoStop}%
\bibitem [{\citenamefont {Coak}\ \emph
  {et~al.}(2019{\natexlab{b}})\citenamefont {Coak}, \citenamefont {Kim},
  \citenamefont {Yi}, \citenamefont {Son}, \citenamefont {Lee},\ and\
  \citenamefont {Park}}]{Coak_2019}%
  \BibitemOpen
  \bibfield  {author} {\bibinfo {author} {\bibfnamefont {M.~J.}\ \bibnamefont
  {Coak}}, \bibinfo {author} {\bibfnamefont {Y.-H.}\ \bibnamefont {Kim}},
  \bibinfo {author} {\bibfnamefont {Y.~S.}\ \bibnamefont {Yi}}, \bibinfo
  {author} {\bibfnamefont {S.}~\bibnamefont {Son}}, \bibinfo {author}
  {\bibfnamefont {S.~K.}\ \bibnamefont {Lee}},\ and\ \bibinfo {author}
  {\bibfnamefont {J.-G.}\ \bibnamefont {Park}},\ }\bibfield  {title} {\bibinfo
  {title} {Electronic and vibrational properties of the two-dimensional {Mott}
  insulator {V}$_{0.9}${PS}$_{3}$ under pressure},\ }\href
  {https://link.aps.org/doi/10.1103/PhysRevB.100.035120} {\bibfield  {journal}
  {\bibinfo  {journal} {Phys. Rev. B}\ }\textbf {\bibinfo {volume} {100}},\
  \bibinfo {pages} {035120} (\bibinfo {year} {2019}{\natexlab{b}})}\BibitemShut
  {NoStop}%
\bibitem [{\citenamefont {Ouvrard}\ \emph {et~al.}(1985)\citenamefont
  {Ouvrard}, \citenamefont {Fréour}, \citenamefont {Brec},\ and\ \citenamefont
  {Rouxel}}]{Coak1985}%
  \BibitemOpen
  \bibfield  {author} {\bibinfo {author} {\bibfnamefont {G.}~\bibnamefont
  {Ouvrard}}, \bibinfo {author} {\bibfnamefont {R.}~\bibnamefont {Fréour}},
  \bibinfo {author} {\bibfnamefont {R.}~\bibnamefont {Brec}},\ and\ \bibinfo
  {author} {\bibfnamefont {J.}~\bibnamefont {Rouxel}},\ }\bibfield  {title}
  {\bibinfo {title} {A mixed valence compound in the two dimensional
  {MPS}$_{3}$ family: {V}$_{0.78}${PS}$_{3}$ structure and physical
  properties},\ }\href
  {https://www.sciencedirect.com/science/article/pii/0025540885902041}
  {\bibfield  {journal} {\bibinfo  {journal} {Mater. Res. Bull.}\ }\textbf
  {\bibinfo {volume} {20}},\ \bibinfo {pages} {1053} (\bibinfo {year}
  {1985})}\BibitemShut {NoStop}%
\bibitem [{\citenamefont {Xue}\ \emph {et~al.}(2019)\citenamefont {Xue},
  \citenamefont {Hou}, \citenamefont {Wang},\ and\ \citenamefont
  {Wu}}]{Xue2019}%
  \BibitemOpen
  \bibfield  {author} {\bibinfo {author} {\bibfnamefont {F.}~\bibnamefont
  {Xue}}, \bibinfo {author} {\bibfnamefont {Y.}~\bibnamefont {Hou}}, \bibinfo
  {author} {\bibfnamefont {Z.}~\bibnamefont {Wang}},\ and\ \bibinfo {author}
  {\bibfnamefont {R.}~\bibnamefont {Wu}},\ }\bibfield  {title} {\bibinfo
  {title} {Two-dimensional ferromagnetic van der {Waals} {CrCl}$_{3}$ monolayer
  with enhanced anisotropy and {Curie} temperature},\ }\href
  {https://link.aps.org/doi/10.1103/PhysRevB.100.224429} {\bibfield  {journal}
  {\bibinfo  {journal} {Phys. Rev. B}\ }\textbf {\bibinfo {volume} {100}},\
  \bibinfo {pages} {224429} (\bibinfo {year} {2019})}\BibitemShut {NoStop}%
\bibitem [{\citenamefont {Kartsev}\ \emph {et~al.}(2020)\citenamefont
  {Kartsev}, \citenamefont {Augustin}, \citenamefont {Evans}, \citenamefont
  {Novoselov},\ and\ \citenamefont {Santos}}]{Kartsev_2020}%
  \BibitemOpen
  \bibfield  {author} {\bibinfo {author} {\bibfnamefont {A.}~\bibnamefont
  {Kartsev}}, \bibinfo {author} {\bibfnamefont {M.}~\bibnamefont {Augustin}},
  \bibinfo {author} {\bibfnamefont {R.~F.~L.}\ \bibnamefont {Evans}}, \bibinfo
  {author} {\bibfnamefont {K.~S.}\ \bibnamefont {Novoselov}},\ and\ \bibinfo
  {author} {\bibfnamefont {E.~J.~G.}\ \bibnamefont {Santos}},\ }\bibfield
  {title} {\bibinfo {title} {Biquadratic exchange interactions in
  two-dimensional magnets},\ }\href
  {https://doi.org/10.1038/s41524-020-00416-1} {\bibfield  {journal} {\bibinfo
  {journal} {npj Comput. Mater.}\ }\textbf {\bibinfo {volume} {6}},\ \bibinfo
  {pages} {150} (\bibinfo {year} {2020})}\BibitemShut {NoStop}%
\end{thebibliography}%

\newpage
\begin{appendix}
	\setcounter{figure}{0}
	\setcounter{table}{0}
	\renewcommand{\thefigure}{S\arabic{figure}}
	\renewcommand{\thetable}{S\arabic{table}}
	\renewcommand{\theequation}{S\arabic{equation}}
	\renewcommand{\tablename}{Table}
	\renewcommand{\figurename}{Fig.}

\titleformat*{\section}{\normalfont\Large\bfseries}
\section*{Supplemental Material for "Contrasting magnetism in VPS$_3$ and CrI$_3$ monolayers with the common honeycomb $S = 3/2$ spin lattice"}

\renewcommand\arraystretch{1.3}
\begin{table}[H]
	\centering
	\caption{The MAE ($\mu$eV/f.u.) calculated using different cutoff energies (eV) and $k$-meshes for VPS$_3$ and CrI$_3$ monolayers with GGA + SOC + $U$.}
	\begin{tabular}{c@{\hskip9mm}c@{\hskip9mm}c@{\hskip9mm}c@{\hskip9mm}}
		\hline\hline
		Systems  &  $E_{\rm cut}$   & $k$-mesh & MAE   \\ 	\hline
		\multirow{3}{*}{VPS$_3$}   & 450 & 9$\times$9$\times$1  &  12 \\
		& 450 & 13$\times$13$\times$1 & 12   \\
		& 550 & 9$\times$9$\times$1   & 12  \\  \hline 
		\multirow{3}{*}{CrI$_3$}   & 450 & 9$\times$9$\times$1   & 503 \\
		& 450 & 13$\times$13$\times$1 & 503  \\
		& 550 & 9$\times$9$\times$1   & 500\\ 
		\hline\hline
	\end{tabular}
	\label{MAE}
\end{table}

\renewcommand\arraystretch{1.3}
\begin{table}[H]
	\centering
	\caption{Relative total energies $\Delta E$ (meV/f.u.) and local spin moments ($\mu_{\rm B}$) for VPS$_3$ monolayer obtained from HSE06 calculations. The derived three exchange parameters (meV) are also listed.}
	\begin{tabular}{c@{\hskip7mm}r@{\hskip7mm}c@{\hskip7mm}r@{\hskip7mm}r@{\hskip7mm}}
		\hline\hline
		System	                        &  States             & $\Delta$\textit{E}  &  V   & S    \\ \hline
		\multirow{4}{*}{VPS$_3$}       & N$\acute{e}$el AF   &  0                  &  $\pm$2.64     & 0.00                          \\
		& stripe AF           &  68                 &  $\pm$2.66     & 0.00                          \\
		& zigzag AF           &  132                &  $\pm$2.69     & 0.00                            \\
		& FM                  &  213                &  2.73   & $-$0.02                                \\
		\multicolumn{5}{c}{\textit{J}$_{1}$ = $-$30.78 \hfill \textit{J}$_{2}$ = $-$0.72 \hfill \textit{J}$_{3}$ = $-$0.78}    \\ 
		\hline\hline
	\end{tabular}
	\label{tbHSE}
\end{table}

\begin{figure}[H]
	\centering 
	\includegraphics[width=9cm]{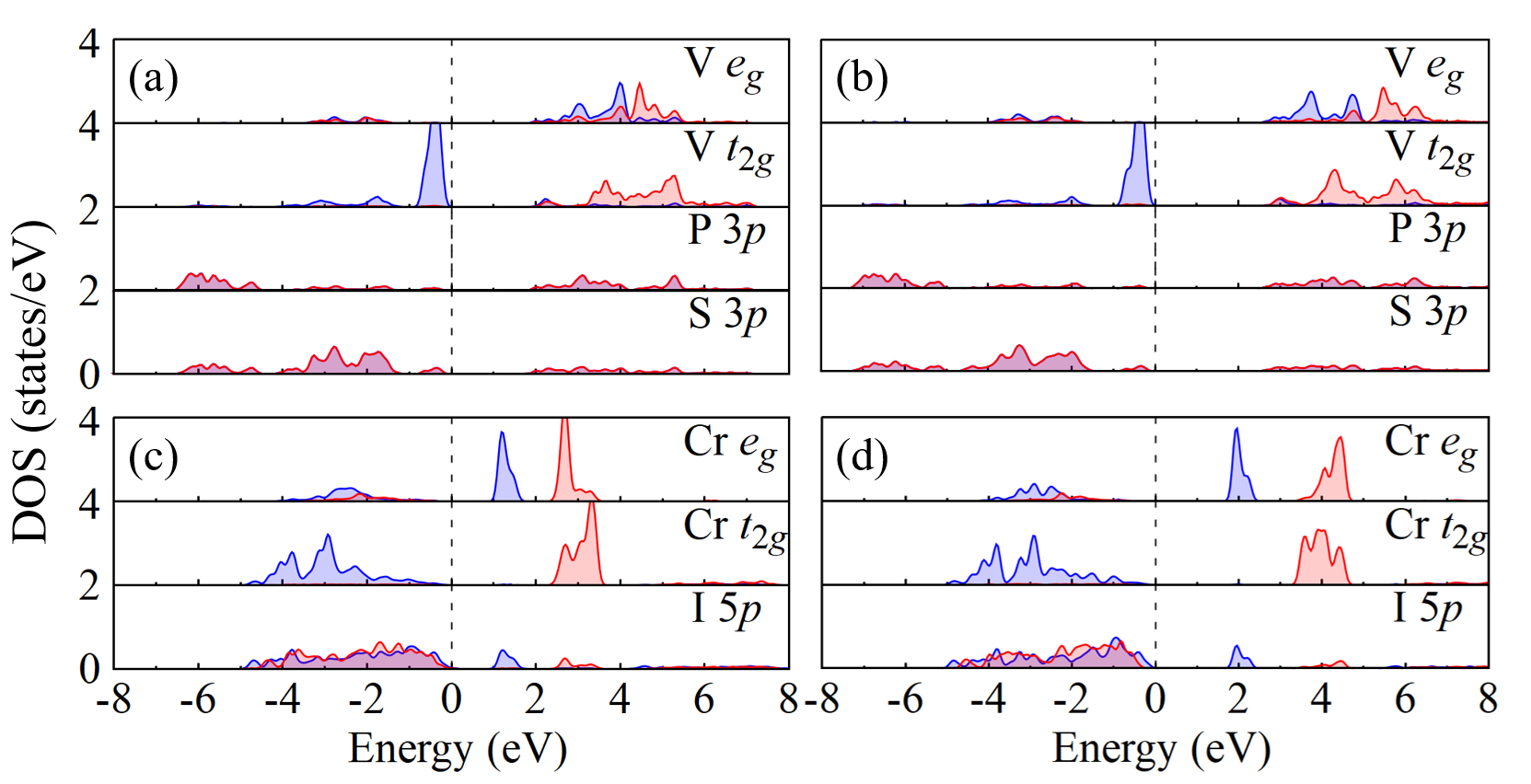}
	\centering
	\caption{Density of states (DOS) of VPS$_3$ monolayer and CrI$_3$ monolayer by (a) and (c) GGA + $U$ calculations, (b) and (d) HSE06 calculations. The Fermi level is set at zero energy. The blue (red) curves stand for the up (down) spin channels.
	}
	\label{MC_K}
\end{figure}
\begin{figure}[H]
	\centering 
	\includegraphics[width=7cm]{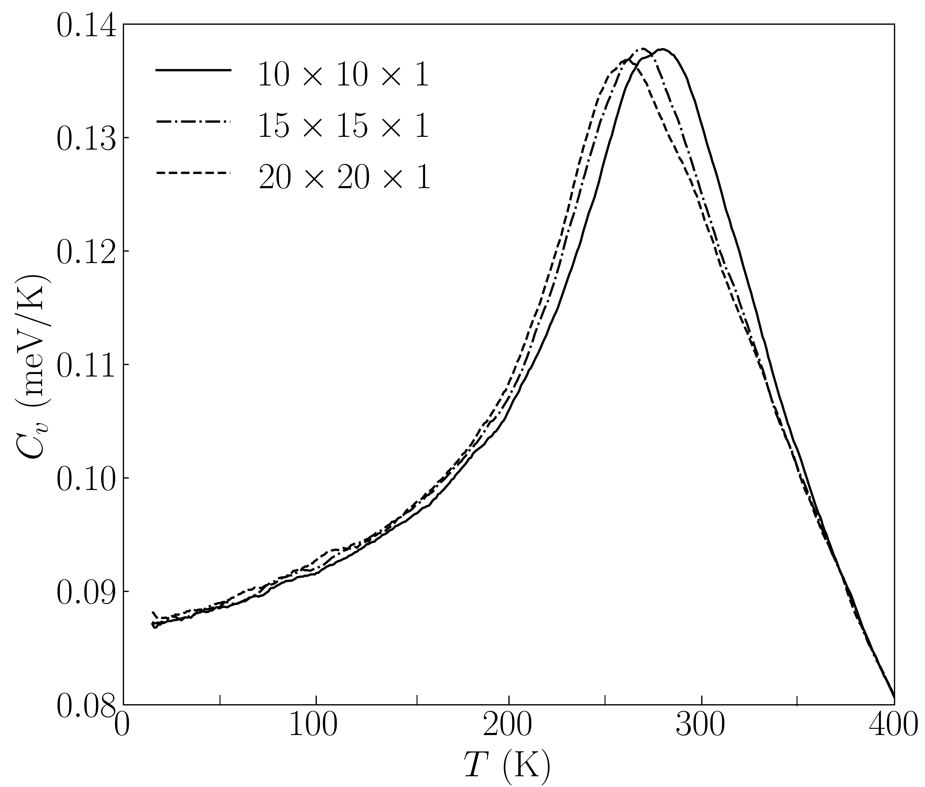}
	\centering
	\caption{Parallel Tempering Monte Carlo (PTMC) simulations of the magnetic specific heat of VPS$_3$ monolayer in the 10$\times$10$\times$1, 15$\times$15$\times$1 and 20$\times$20$\times$1 lattices.
	}
	\label{MC_K}
\end{figure}

\begin{figure}[H]
	\includegraphics[width=8.5cm]{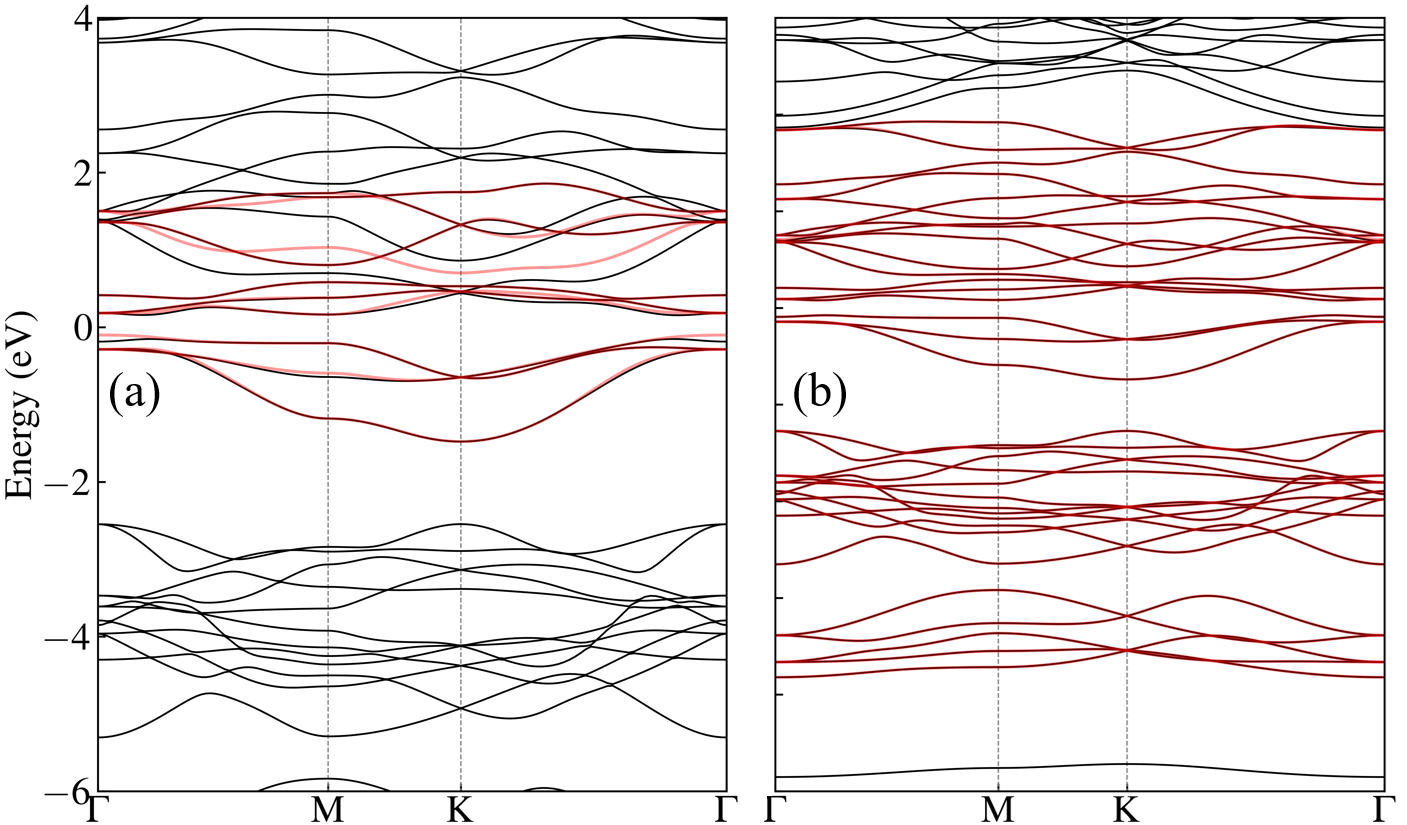}
	\centering
	\caption{The calculated DFT (black solid lines) and Wannier-interpolated (red dashed lines) band structures of (a) projected onto V 3$d$ orbitals and (b) simultaneously projected onto V 3$d$, S 3$p$, and P 3$p$ orbitals for the VPS$_3$ monolayer.}
\end{figure}

\begin{figure}[H]
	\includegraphics[width=8cm]{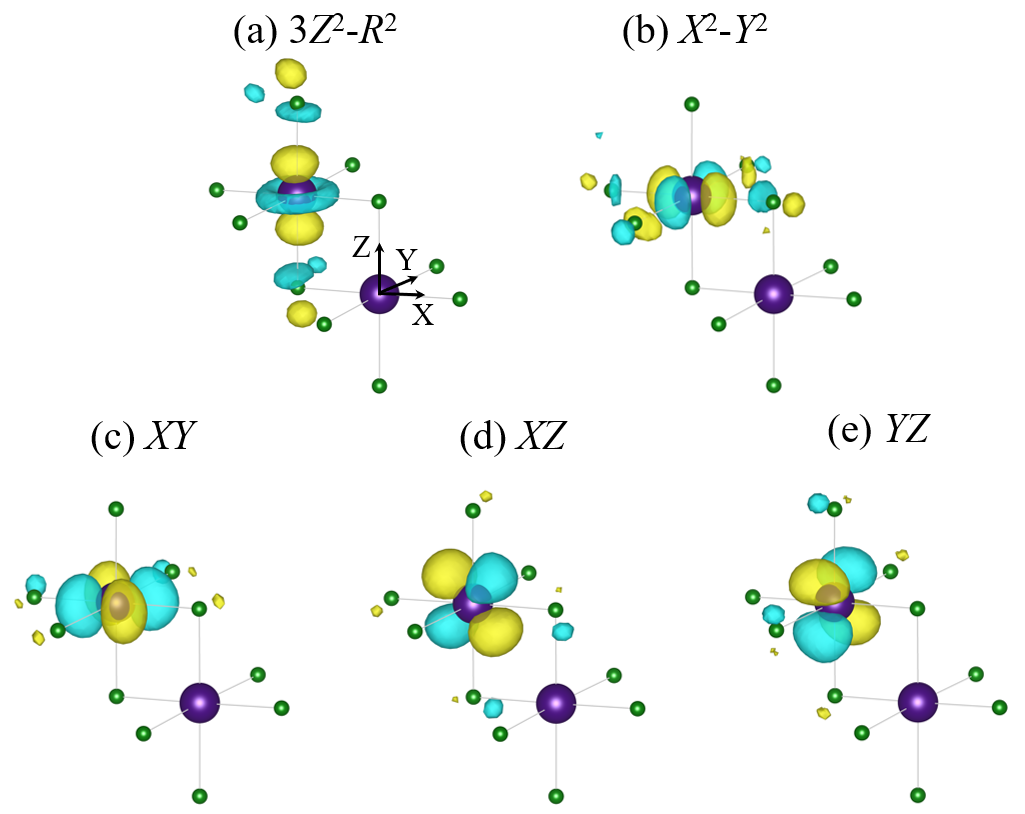}
	\centering
	\caption{(a)-(e) Contour-surface plots of the V 3$d$ Wannier functions projected onto V 3$d$ orbitals for the VPS$_3$ monolayer. For all plots, an isosurface level of $\pm$3.0 is chosen (yellow for positive values and cyan for negative values), using the VESTA visualization program.}
\end{figure}

\begin{figure}[H]
	\centering 
	\includegraphics[width=8.5cm]{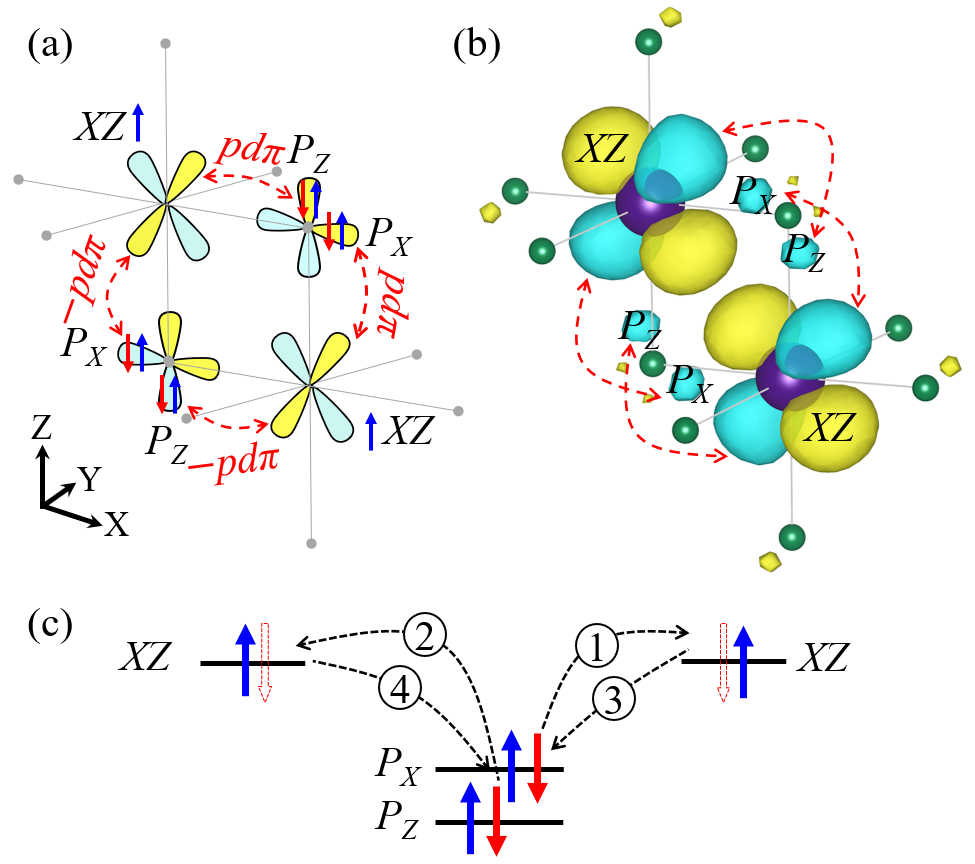}
	\centering
	\caption{(a) The hybridization between the S $p_Z$ orbital and the $XZ$ orbital,  and between the S $p_X$ and the $XZ$ in VPS$_3$. (b) The corresponding Wannier orbitals. (c) The indirect ($XZ$)-($p_X$, $p_Z$)-($XZ$) hopping channels lead to weak FM coupling.}
\end{figure}

\begin{figure}[H]
	\includegraphics[width=8cm]{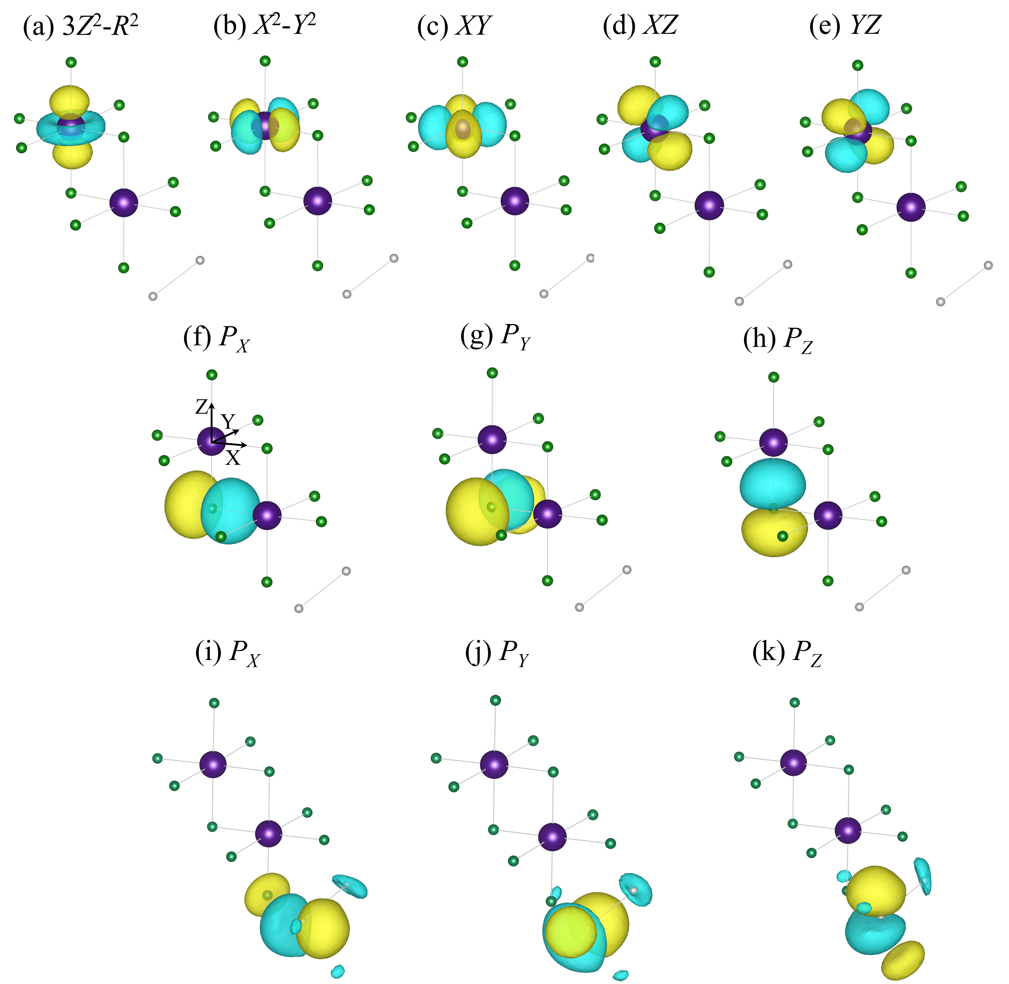}
	\centering
	\caption{Contour-surface plots of the (a)-(e) V 3$d$, (f)-(h) S 3$p$ and (i)-(k) P 3$p$ Wannier functions simultaneously projected onto V 3$d$, S 3$p$, and P 3$p$ orbitals for the VPS$_3$ monolayer. For all plots, an isosurface level of $\pm$3.0 is chosen (yellow for positive values and cyan for negative values) using the VESTA visualization program.}
\end{figure}

\begin{figure}[H]
	\centering 
	\includegraphics[width=8.5cm]{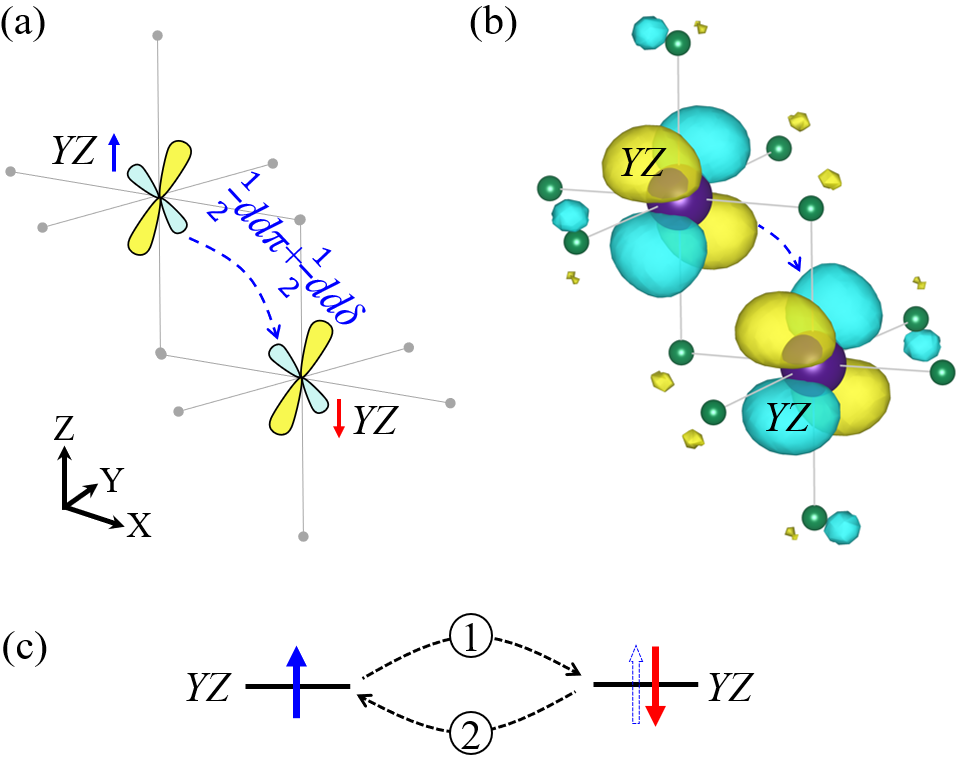}
	\centering
	\caption{(a) The hybridization between the two $YZ$ orbitals and (b) the corresponding Wannier orbitals in VPS$_3$. (c) The direct $d$-$d$ hopping channels between two $YZ$ orbitals lead to AF coupling.}
\end{figure}
\begin{figure}[H]
	\centering 
	\includegraphics[width=8.5cm]{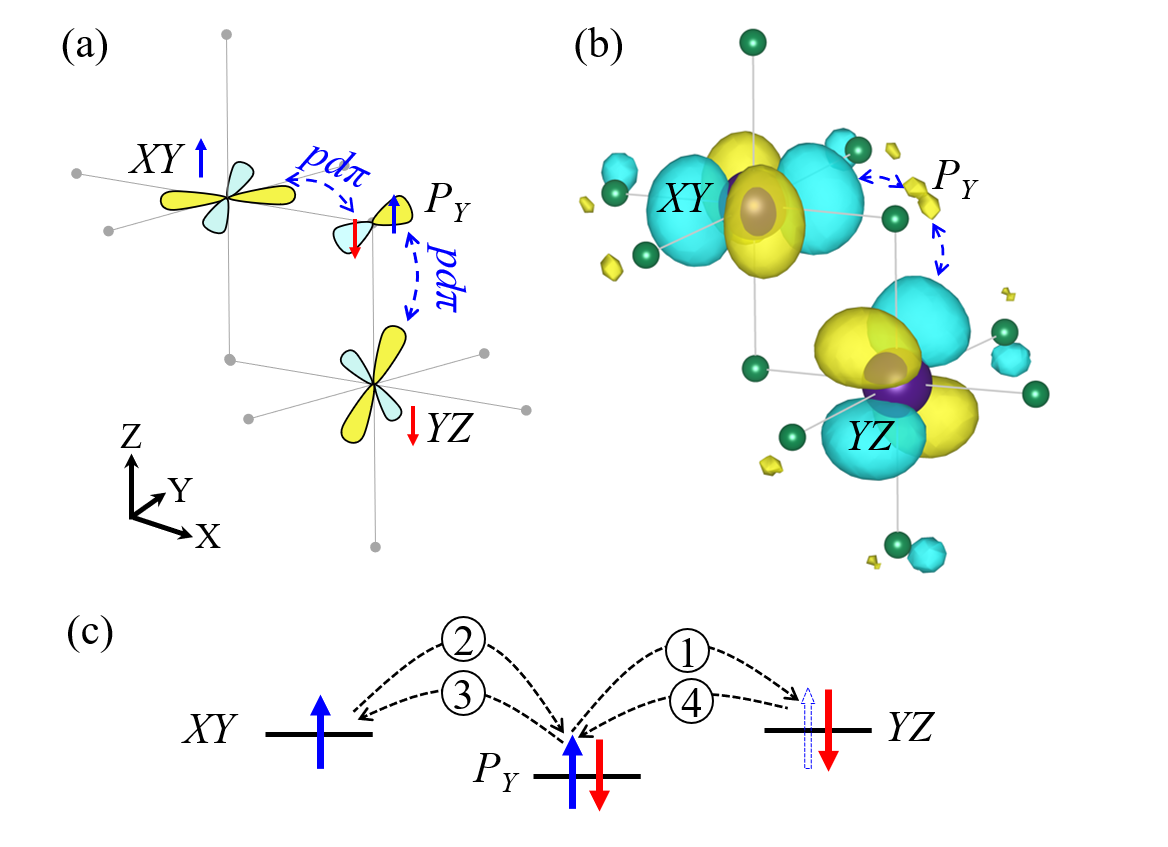}
	\centering
	\caption{(a) The hybridization between the S $p_Y$ orbital and the $XY$ orbital, as well as between the S $p_Y$ orbital and the $YZ$ orbital in VPS$_3$. (b) The corresponding Wannier orbitals. (c) The indirect ($XY$)-($p_Y$)-($YZ$) hopping channels lead to AF coupling.}
\end{figure}
\renewcommand\arraystretch{1.3}
\begin{table}[H]
	\centering
	\caption{The hopping parameters (meV) obtained from Wannier functions of VPS$_3$ monolayer, which are simultaneously projected onto V 3$d$, S 3$p$, and P 3$p$ orbitals. By referring to Fig. 4(a) in the main text, the S-$p_Z$ ($p_X$) orbital has a normal $pd\pi$ hopping of 619 meV with one V (the other V$'$) $d_{XZ}$ orbital , but the S-$p_Z$ ($p_X$) has a tiny hopping of 28-30 meV--less than 5\% of the normal $pd\pi$ hopping--with the other V$'$ (one V) $d_{XZ}$ orbital due to the small deviation of the V-S-V$'$ bond angle from the ideal 90 degrees (otherwise, this hopping is exactly zero due to the symmetry restricted orthogonality).
	}
	\begin{tabular}{c@{\hskip9mm}l@{\hskip9mm}r@{\hskip9mm}r@{\hskip9mm}r@{\hskip9mm}} \hline\hline
		\multicolumn{2}{c}{\multirow{2}{*}{Hopping (\textit{t})}}  & \multicolumn{3}{c}{S} \\		      
		& & $P_X$ & $P_Y$ & $P_Z$ \\ \hline
		\multirow{5}{*}{V} & $3Z^2-R^2$ & --670 & --22 & --18  \\
		& $X^2-Y^2$ & 1055 & 103 & --27  \\
		& $XY$ & 95 & --582 & --37   \\
		& $XZ$ & 30 & --19 & --619  \\
		& $YZ$ & --48 & 61 & --40  \\  \hline 
		\multirow{5}{*}{V$'$} & $3Z^2-R^2$ & --16 & 100 & 1249  \\
		& $X^2-Y^2$ & --29 & 32 & --53  \\
		& $XY$ & --40 & 61 & --48  \\
		& $XZ$ & --619 & --19 & 28  \\
		& $YZ$ & --38 & --582 & 94  \\ 
		\hline\hline
	\end{tabular}
	\label{tbhopping}
\end{table}
\renewcommand\arraystretch{1.3}
\begin{table}[H]
	\centering
	\caption{The hopping parameters (meV) obtained from Wannier functions of VPS$_3$ monolayer, which are simultaneously projected onto V 3$d$, S 3$p$, and P 3$p$ orbitals.}
	\begin{tabular}{c@{\hskip3mm}l@{\hskip3mm}r@{\hskip3mm}r@{\hskip3mm}r@{\hskip3mm}r@{\hskip3mm}r@{\hskip3mm}} \hline\hline
		\multicolumn{2}{c}{\multirow{2}{*}{Hopping (\textit{t})}}  & \multicolumn{5}{c}{V$_0$} \\		      
		& & $3Z^2-R^2$ & $X^2-Y^2$ & $XY$ & $XZ$ & $YZ$ \\ \hline
		\multirow{5}{*}{V$_1$} & $3Z^2-R^2$ & --132 & 109 & --4 & 19 & --9 \\
		& $X^2-Y^2$ & 109 & --7 & --12 & 32 & --9 \\
		& $XY$ & --4 & --12 & 69 & --14 & --56 \\
		& $XZ$ & 19 & 32 & --14 & --341 & --15 \\
		& $YZ$ & --9 & --9 & --56 & --15 & 69 \\  \hline 
		\multirow{5}{*}{V$_2$} & $3Z^2-R^2$ & --22 & 16 & 20 & 5 & 8 \\
		& $X^2-Y^2$ & --16 & 5 & --3 & 5 & 7 \\
		& $XY$ & 20 & 3 & --3 & --6 & 8 \\
		& $XZ$ & 8 & --7 & 8 & --14 & 7 \\
		& $YZ$ & 5 & --5 & --6 & 15 & --14 \\ \hline
		\multirow{5}{*}{V$_3$} & $3Z^2-R^2$ & 15 & 17 & 1 & 1 & --4 \\
		& $X^2-Y^2$ & 17 & --4 & --2 & --2 & 7 \\
		& $XY$ & 1 & --2 & --2 & 7 & --17 \\
		& $XZ$ & 1 & --2 & 7 & --2 & --17 \\
		& $YZ$ & --4 & 7 & --17 & --17 & 67 \\ 
		\hline\hline
	\end{tabular}
	\label{tbhopping}
\end{table}
\begin{figure}[H]
	\centering 
	\includegraphics[width=8.5cm]{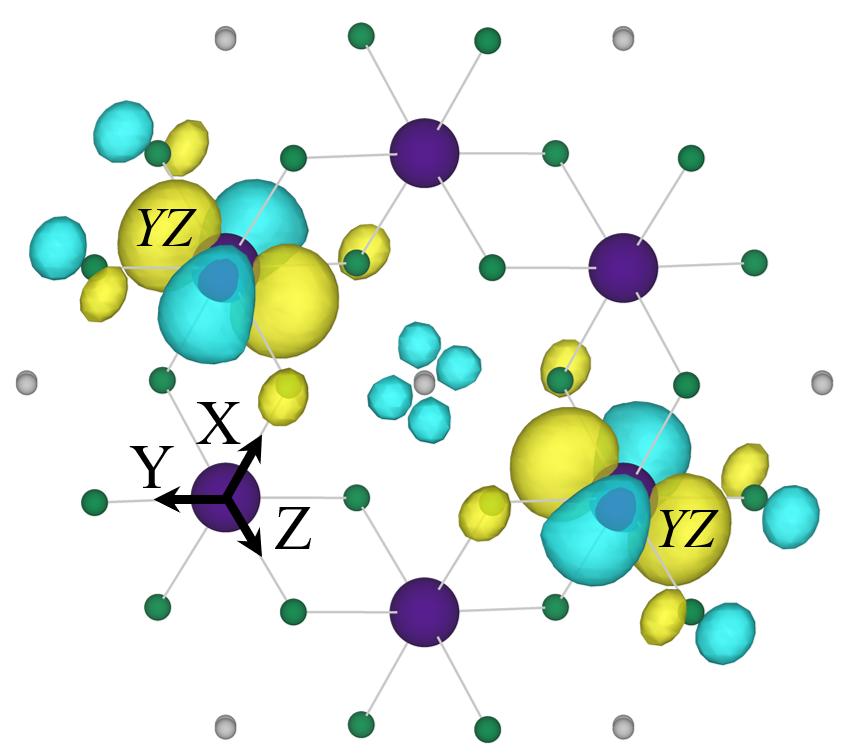}
	\centering
	\caption{The 3NN Wannier orbitals between the two $YZ$ orbtials in VPS$_3$.}
\end{figure}
\begin{figure}[H]
	\centering 
	\includegraphics[width=9cm]{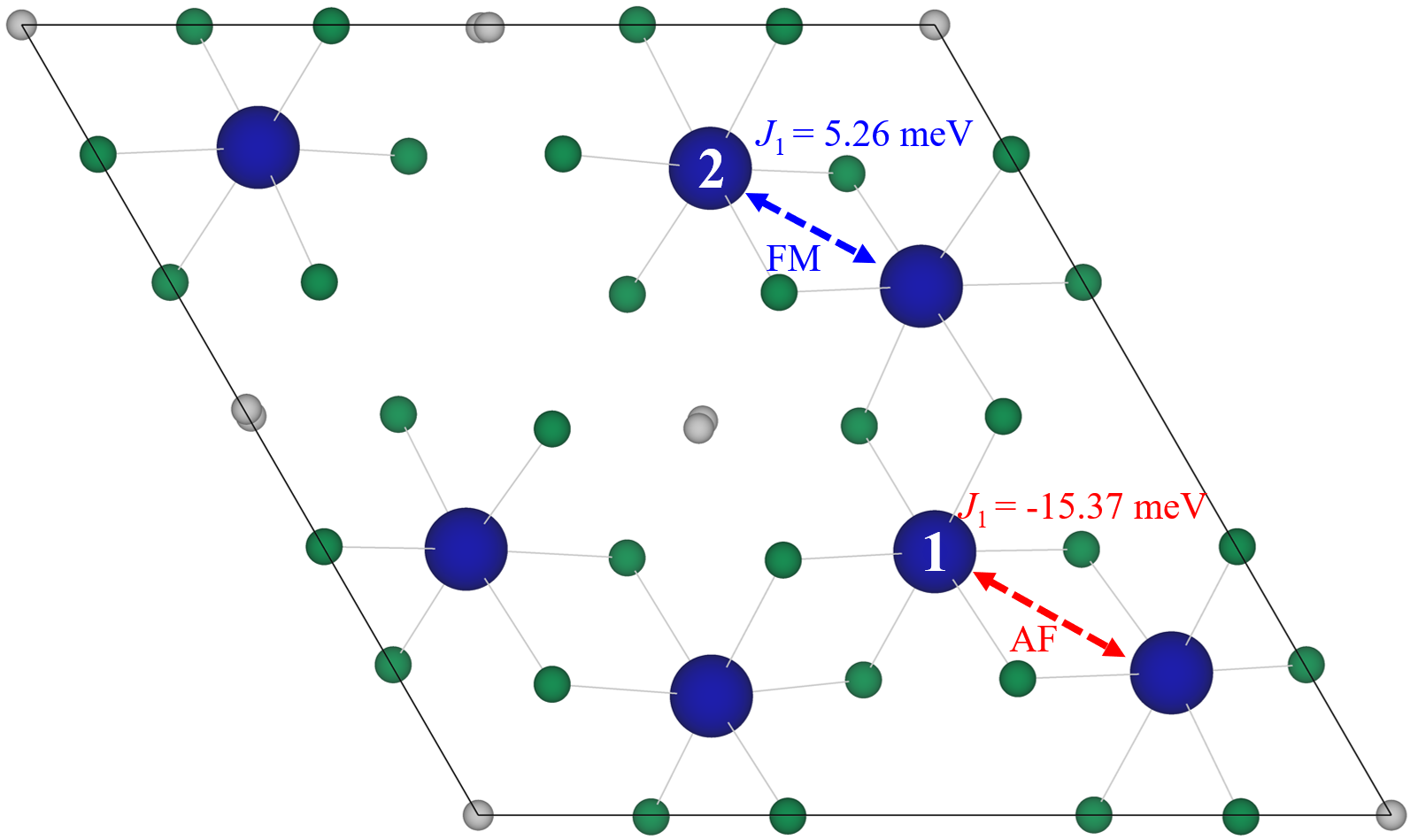}
	\centering
	\caption{The 2$\times$2 supercell of VPS$_3$ with a single V vacancy  (with the vacancy ratio of 1/8). Our GGA + $U$ calculations show: For the V$_1$ site farthest from the vacancy, the averaged first-nearest-neighbor (1NN) AF exchange parameter is $-15.37$ meV, estimated from the energy difference of $-$207.47 meV (3$JS^2$ for AF $vs$ $-3JS^2$ for FM, both in 3-fold 1NN coordination, $J = -$207.47 meV/$6S^2$) when only the V$_1$ spin is flipped; For the V$_2$ site nearest to the vacancy, the averaged 1NN FM exchange parameter is 5.26 meV, estimated from the energy difference of 47.34 meV ($2JS^2$ for AF $vs$ $-2JS^2$ for FM, both in 2-fold 1NN coordination due to the 1NN V vacancy, $J$ = 47.34 meV/$4S^2$) when only the V$_2$ spin is flipped.}
	\label{vacancy}
\end{figure}

\begin{figure}[H]
	\includegraphics[width=8.5cm]{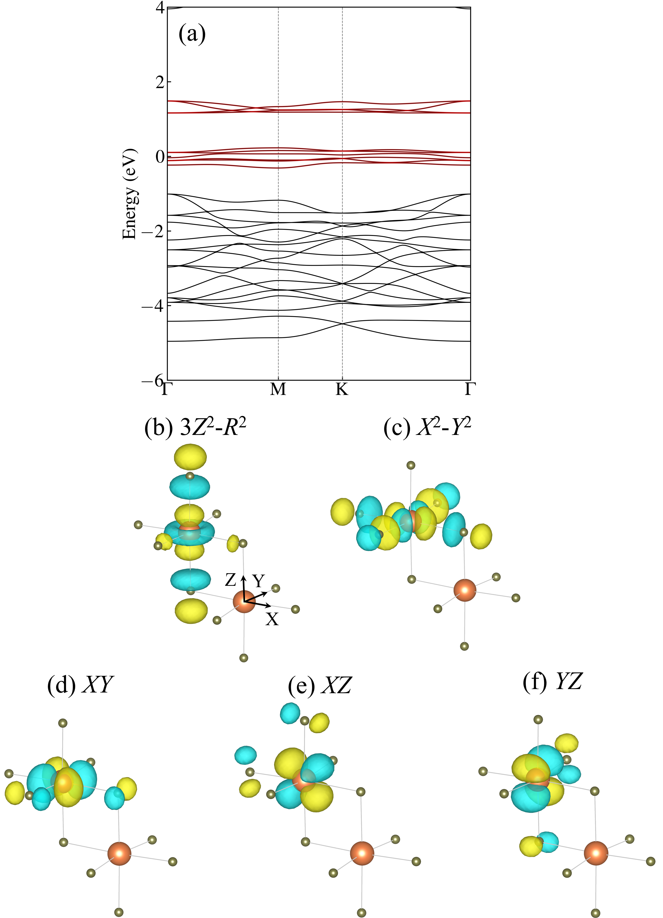}
	\centering
	\caption{(a) The calculated DFT (black solid lines) and Wannier-interpolated (red dashed lines) band structures projected onto Cr 3$d$ orbitals for the CrI$_3$ monolayer. (b)-(f) Contour-surface plots of the Cr 3$d$ Wannier functions. For all plots, an isosurface level of $\pm$3.0 (yellow for positive values and cyan for negative values) is chosen using the VESTA visualization program.}
\end{figure}

\begin{figure}[H]
	\centering 
	\includegraphics[width=8.5cm]{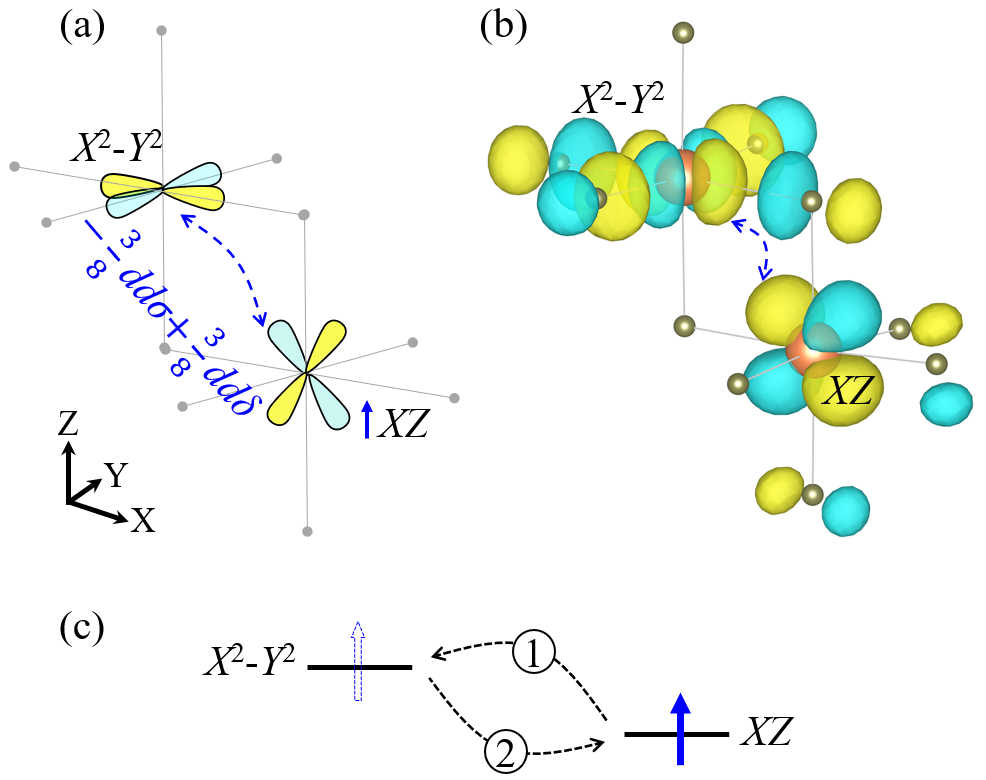}
	\centering
	\caption{(a) The hybridization between the $X^2-Y^2$ and $XZ$ orbitals and (b) the corresponding Wannier orbitals in CrI$_3$. (c) The direct $d$-$d$ hopping channels between the $X^2-Y^2$ and $XZ$ orbitals lead to FM coupling.}
\end{figure}
%


\end{appendix}

\end{document}